\documentclass[letterpaper]{article} 
\usepackage{aaai24}  
\usepackage{times}  
\usepackage{helvet}  
\usepackage{courier}  
\usepackage[hyphens]{url}  
\usepackage{graphicx} 
\usepackage{subcaption}
\usepackage{amsmath}
\usepackage{amssymb}
\usepackage{amsthm}  
\usepackage[english]{babel}
\usepackage{float}
\usepackage{times}
\usepackage{helvet}
\usepackage{courier}
\usepackage{multirow}
\usepackage{booktabs}
\usepackage{amsfonts}
\usepackage{cleveref}
\usepackage{caption}
\usepackage{array}
\usepackage{makecell}
\usepackage{float}
\usepackage{xcolor}
\usepackage{lipsum}
\usepackage{mathtools}

\DeclareMathOperator*{\argmax}{argmax}
\def\UrlFont{\rm}  
\usepackage{natbib}  
\usepackage{caption} 
\usepackage{algorithm}
\usepackage{algorithmic}
\usepackage{multirow}
\usepackage[utf8]{inputenc}
\usepackage{aaai24}
\usepackage{newfloat}
\usepackage{listings}
\DeclareCaptionStyle{ruled}{labelfont=normalfont,labelsep=colon,strut=off} 
\floatstyle{ruled}
\newfloat{listing}{tb}{lst}{}
\floatname{listing}{Listing}
\usepackage{xcolor}

\title{EarnHFT: Efficient Hierarchical Reinforcement Learning for \\ High Frequency Trading}
\author {
    Molei Qin\equalcontrib,
    Shuo Sun\equalcontrib,
    Wentao Zhang,
    Haochong Xia,
    Xinrun Wang\textsuperscript{$\dagger$},
    Bo An\textsuperscript{$\dagger$},
}
\affiliations {
    Nanyang Technological University, Singapore\\
    \{haochong001, shuo003\}@e.ntu.edu.sg,
    \{molei.qin, wt.zhang, xinrun.wang, boan\}@ntu.edu.sg
}

\begin{document}

\maketitle

\begin{abstract}
High-frequency trading (HFT) uses computer algorithms to make trading decisions in short time scales (e.g., second-level), which is widely used in the Cryptocurrency (Crypto) market (e.g., Bitcoin). 
Reinforcement learning (RL) in financial research has shown stellar performance on many quantitative trading tasks. However, most methods focus on low-frequency trading, e.g., day-level, which cannot be directly applied to HFT because of two challenges. First, RL for HFT involves dealing with extremely long trajectories (e.g., 2.4 million steps per month), which is hard to optimize and evaluate. Second, the dramatic price fluctuations and market trend changes of Crypto make existing algorithms fail to maintain satisfactory performance.
To tackle these challenges, we propose an \textbf{E}fficient hie\textbf{A}rchical \textbf{R}einforcement lear\textbf{N}ing method for \textbf{H}igh \textbf{F}requency \textbf{T}rading (EarnHFT), a novel three-stage hierarchical RL framework for HFT. In stage I, we compute a Q-teacher, i.e., the optimal action value based on dynamic programming, for enhancing the performance and training efficiency of second-level RL agents.
In stage II, we construct a pool of diverse RL agents for different market trends, distinguished by return rates, where hundreds of RL agents are trained with different preferences of return rates and only a tiny fraction of them will be selected into the pool based on their profitability.
In stage III, we train a minute-level router which dynamically picks a second-level agent from the pool to achieve stable performance across different markets.
Through extensive experiments in various market trends on Crypto markets in a high-fidelity simulation trading environment, we demonstrate that EarnHFT significantly outperforms 6 state-of-art baselines in 6 popular financial criteria, exceeding the runner-up by 30\% in profitability.
\end{abstract}

\section{Introduction}

High-frequency trading (HFT), taking up more than 73\% volume in the fincial market, refers to leveraging complicated computer algorithms or mathematical models to place or cancel orders at incredibly short time scales~\cite{ALMEIDA2023100785}. A good HFT strategy enables investors to make more profit than a low-frequency strategy and is therefore pursued by many radical traders. It has been widely used in Cryptocurrency (Crypto) market due to Crypto's 24/7 non-stop trading time, which prevents Crypto holders from overnight risk, and dramatic price fluctuations, which provides more profitable trading opportunities for HFT.

Although reinforcement learning (RL) algorithms~\cite{sun2022deepscalper,theate2021application,cumming2015investigation} have achieved outstanding results in low-frequency trading in traditional financial markets like stock or futures, few maintain robust performance under the setting of HFT due to two challenges:
\begin{itemize} 
    \item An extremely large time horizon induces low data efficiency for RL training. Compared with Atari games where the time horizon is 6000~\cite{mnih2013playing}, the time horizon of HFT is around 1 million, because second-level agents need to be evaluated in dozens of days.\footnote{$60 \text{s/m} \times 60 \text{m/h}\times 24\text{h/d} \times 12 d=1036800\text{s}$} Large time horizons need more data to converge~\cite{zhang2023towards}, demanding more computational resources.
    \item The dramatic market changes cause agents trained on history data to fail in maintaining performance over long periods. In a traditional RL setting, the training and testing environments remain consistent. However, Crypto market trend changes cause a significant difference between the training and testing environments. An agent trained on one market trend tends to cause tremendous losses once the trend changes dramatically in the market.
\end{itemize}
To tackle the challenges, we propose an \textbf{E}fficient hie\textbf{A}rchical \textbf{R}einforcement lear\textbf{N}ing method for \textbf{H}igh \textbf{F}requency \textbf{T}rading (EarnHFT) as shown in Figure~\ref{fig:pipline}. 
In stage I, we build a Q-teacher indicating the optimal action value based on dynamic programming and future price information, which is used as a regularizer to train RL agents delivering a target position every second for better performance and faster training speed. 
In stage II, we first train hundreds of second-level RL agents following the stage I process under different market trend preferences, where buy and hold~\cite{shiryaev2008thou} return rates are used as the preference indicators. We further label each market based on DTW~\cite{muda2010voice} as different categories and use the profitability performance under each market category to select a tiny fraction of trained second-level RL agents to construct a strategy pool.  
In stage III, we train a router which dynamically picks a second-level agent from the pool per minute to achieve stable performance across different markets.
Through extensive experiments in various market trends in the Crypto market under a high-fidelity simulation trading environment, we demonstrate that EarnHFT significantly outperforms 6 state-of-art baselines in terms of 6 popular financial criteria, exceeds the runner-up by at least 30\% in terms of profitability.
\section{Related Works}
In this section, we introduce the related works on traditional finance methods used in HFT and RL for quantitative trading. More discussion can be found in Appendix A.
\subsection{High-Frequency Trading in Crypto}
HFT, aiming to profit from slight price fluctuation in a short period of time in the market, has been widely used in companies~\cite{zhou2021high}. Crypto traders invest differently from those in stock markets because of the high volatility~\cite{delfabbro2021psychology}.
In the Crypto market, there are many high-frequency technical indicators~\cite{huang2019predicting}, such as imbalance volume (IV)~\cite{chordia2002order} and moving average convergence divergence (MACD)~\cite{krug2022enforcing}, to capture buying and selling pressures among different time scales.
However, these technical indicators also have limitations. In the volatile Crypto market, technical indicators may produce false signals. The result is sensitive to hyper-parameter settings, e.g. the take profit point and the stop loss.
\subsection{RL for Quantitative Trading}
Many deep reinforcement learning methods for quantitative trading have been proposed. DeepScalper~\cite{sun2022deepscalper} uses a hindsight bonus and auxiliary task to improve the model's generalization ability in intraday trading. DRA~\cite{briola2021deep} uses LSTM and PPO. CDQNRP~\cite{zhu2022quantitative} uses a random perturbation to increase the stability of training a convolution DQN. 
However, these algorithms focus mainly on designing only one powerful RL agent to conduct profitable trading in short-term scenarios, neglecting its failure of maintaining performance over long periods.

Hierarchical Reinforcement Learning (HRL), which decomposes a long-horizon task into a hierarchy of sub-problems, has been studied for decades. 
There are some hierarchical RL frameworks for quantitative trading. HRPM~\cite{wang2021commission} utilizes a hierarchical framework to simulate portfolio management and order execution. MetaTrader~\cite{niu2022metatrader} proposes a router to pick the most suitable strategy for the current market situation. 
However, these hierarchical frameworks are all utilized in portfolio management. Its application remains unexplored in HFT where only one asset is traded.

\section{Problem Formulation}
In this section, we present some basic finance concepts used in simulating the trading process and propose a hierarchical Markov decision process (MDP) framework for HFT\footnote{More detailed discussions are described in Appendix B.}.
\subsection{Financial Foundations for HFT}
We first introduce some basic financial concepts used to describe state, reward, and action in the following hierarchical Markov Decision Process (MDP) framework and present the objective of HFT.

\noindent
\textbf{Limit Order Book} (LOB) records unfilled orders. It is widely used to describe the market micro-structure~\cite{madhavan2000market} in finance. We denote an m-level LOB at time $t$ as $b_t=(p_t^{b_1},p_t^{a_1},q_t^{b_1},q_t^{a_1},...,p_t^{b_m},p_t^{a_m},q_t^{b_m},q_t^{a_m})$, where $p_t^{b_i}$  ($p_t^{a_i}$)
is the level $i$ bid (ask) price, $q_t^{b_i}$ ($q_t^{a_i}$) is the quantity.

\noindent
\textbf{OHLC} is aggregated information of executed orders. OHLC vector at time $t$ is
denoted as $x_t=(p_t^o,p_t^h,p_t^l,p_t^c)$, where $p_t^o,p_t^h,p_t^l,p_t^c$ indicate the open, high, low and close price.

\noindent
\textbf{Technical Indicators} indicate features calculated by a formulaic combination of the original OHLC or LOB to uncover the underlying pattern of the financial market. We denote the technical indicator vector at time $t$ $y_t=\phi (b_t,x_t,...,b_{t-h},x_{t-h})$, where $\phi$ is the function that maps OHLC and LOB to technical indicators.

\noindent
\textbf{Market Order} is a trade to buy or sell a financial asset instantly. The executed price is calculated as Equation~\ref{equation:calculate_market_order}.
\begin{equation}
\label{equation:calculate_market_order}
E_t(M) = \sum_{i} (p_t^i \times \min(q_t^i, R_{i-1}))(1+\sigma)
\end{equation}
where $E$ is the execution price, $R_{i-1}$ is the remaining quantity after level $i$ in LOB, $\sigma$ is the commission fee rate, and $q_t^i,p_t^i$ are the level $i$ price and quantity in LOB respectively. 

\noindent
\textbf{Position} is the amount of a financial asset traders hold. Position at time $t$ is denoted as $P_t$ and $P_t \geq 0$, indicating only a long position is permitted in this formulation.

\noindent
\textbf{Net Value} $V_t$ is the sum of cash and value of the position, calculated as $V_t=V_{ct}+P_t\times p_t^{b1}$, where $V_{ct}$ is the cash.

We aim to maximize the net value by conducting market orders on a single asset based on market information (e.g., LOB and OHLC) at a second-level time scale. 
\begin{figure*}[th]
\begin{center}
\includegraphics[width=\textwidth]{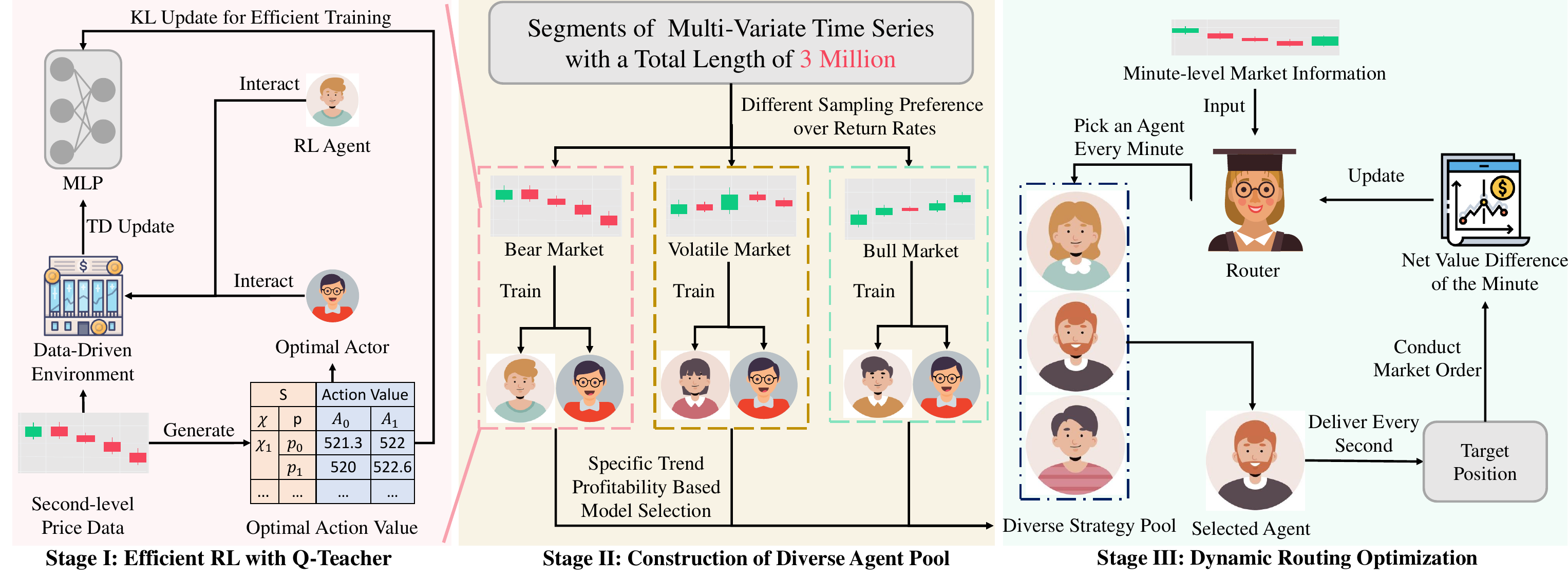}
\end{center}
\caption{The overview of EarnHFT. First, we compute a Q-teacher for enhancing the performance and training efficiency of second-level RL agents.
Then, we efficiently train diverse RL agents under various market trends and select a small fraction of them to form an agent pool based on profitability.
Finally, we train a minute-level router which dynamically picks a second-level agent from the pool to achieve stable performance across different markets.}
\label{fig:pipline}
\end{figure*}

\subsection{Hierarchical MDP Framework}
In this subsection, we formulate HFT as a hierarchical MDP. An MDP is defined by the tuple: ($S$, $A$, $P$, $r$, $\gamma$, $T$), where $S$ is the state space and $A$ is the action space. $P: S\times A\times S \rightarrow [0, 1]$ is the transition function, $r: S\times A\times S\rightarrow R$ is the reward function, $\gamma \in (0, 1]$ is the discount factor and $T$ is the time horizon. In an MDP, the agent receives the current state $s_t \in S$ from the environment, performs an action $a_{t} \in A$, and gets the next state $s_{t+1} \in S$ and a reward $r_t$. An agent’s policy is defined by $\pi_\theta$ : $S \times A \rightarrow [0, 1]$, which is parameterized by $\theta$. The objective of the agent is to learn an optimal policy $\pi^*=\arg\max_\theta E_{\pi_\theta}[\sum_{t=0}^T \gamma^tr_t|S_0]$ where $s_0$ is the initial state of the MDP.

In RL for HFT, data drifting of the micro-level market information prevents a single agent from maintaining its performance over long periods and it is difficult to train a profitable agent under all trends because of the conflict in effective strategies under different market conditions. Macro-level information, as an aggregation of micro-level market information, provides insight into the dynamics of the micro-level market.
Therefore, we formulate HFT as a hierarchical MDP, where the low-level MDP operating on a second-level time scale formulates the process of micro-level market dynamics and trading execution and the high-level MDP operating on a minute-level time scale formulates the process of the macro-level market trends and strategy adjustment. It is defined by ($\text{MDP}_h$, $\text{MDP}_l$), where
\begin{align*}
   \text{MDP}_h &= (S_h, A_h, P_h, R_h, \gamma_h, T_h)\\
   \text{MDP}_l &= (S_l, A_l, P_l, R_l, \gamma_l, T_l)
\end{align*}
\noindent
\textbf{Low-level State} $S_{lt}$ at time $t$ consists of two parts: the latent representation $X_{lt}$, which is the micro-level market technical indicators and private state $P_t$. 
$X_{lt}$ consists of 54 features and is calculated as $X_{lt}=\phi_l(C_{lt})$ where $C_{lt}$ is a rolling window of second-level OHLC and the snapshot of LOB with length 60. $P_t$ indicates the current position of the agent.

\noindent
\textbf{Low-level Action} $a_{lt}$ at time $t$ is the target position. It is chosen from a predefined position pool $A_l$ with finite elements defined as $\{0, \frac{H}{|A|-1},..., H\}$ where $|A|$ represents the number of the action choices and $H$ is the maximum position. If $a_{lt} \neq P_t$, then we instantly take a market order $E_t(a_{lt}-P_t)$, making the current position to $a_{lt}$.

\noindent
\textbf{Low-level Reward} $r_{lt}$ at time $t$ is the net value differential in the second-level time scale, referring to money made through one second. It is calculated as $r_{lt}=P_{t+1}\times p_{t+1}^{b1}-(P_t\times p_t^{b1}+E_t(P_{t+1}-P_t)$, where we use the best bid price to calculate the value of the current position.

\noindent
\textbf{High-level State} $S_{ht}$ at time $t$ also consists of two parts: the latent representation $X_{ht}$, which is the macro-level market technical indicators and private state $P_t$. $X_{ht}$ consists of 19 features and is calculated as $X_{ht}=\phi_h(C_{ht})$ where $C_{ht}$ is a rolling window of minute-level OHLC length 60. $P_t$ indicates the current position of the agent.

\noindent
\textbf{High-level Action} $a_{ht}$ is the selected agent at time $t$. It is chosen from a pre-trained agent pool $A_h$, each of which is trained under a low-level MDP. 

\noindent
\textbf{High-level Reward} $r_{lt}$ at time $t$ is the net value differential in the minute-level time scale, referring to money made through one minute. It is also the return of the selected low-level agent makes under low-level MDP in one minute and is calculated as $r_{hT}=\sum_{t=T}^{T+\tau}r_{lt}$.

In this bilevel hierarchical MDP framework, for every minute, our high-level agent picks a low-level agent, which will adjust its position every second to make profit. We aim to find a set of low-level agents (traders) and a high-level agent (router) to maximize our total profit.

\section{EarnHFT}
In this section, we demonstrate three stages of EarnHFT as shown in Figure~\ref{fig:pipline}. In stage I, we present RL with Q-teacher, which improves the training efficiency, to train low-level agents. In stage II, agents are trained and evaluated in different market trends, forming a diverse pool for hierarchical constructions. In stage III, we train a router to pick a proper agent to maintain profitability in the non-stationary market.
\subsection{Stage I: Efficient RL with Q-Teacher}
A long trajectory causes extra computational cost in traditional RL settings. However, in our low-level MDP, the price information is not influenced by our policy. By using future price information and dynamic programming, we can easily construct the optimal action value~\cite{sutton2018reinforcement} to help train RL agents more efficiently. Here, we use an optimal value supervisor and an optimal actor to aid training.

\noindent
\textbf{Optimal Value Supervisor.} 
Although using RL to conduct HFT suffers from drawbacks such as overfitting stated in ~\cite{zhang2023generalizable}, it can compute the optimal action value for any state, unlike traditional RL where even expert trajectories are hard to acquire.
Since our position choice is finite, we can backward calculate the optimal action value. By calculating the market order costing and the value of position fluctuations, we can get the reward and calculate the action value for the previous state as shown in Algorithm~\ref{alg:q_table}.
Adding optimal action value as a supervision signal can help the agent to explore faster and get positive rewards more quickly.
During the training of the DDQN~\cite{van2016deep} agent, we can add a supervision term which is the Kullback–Leibler (KL) divergence between the agent's action values and the optimal action value picked from the same state. 
Let $Q_t(\chi, p, a)$ denote the action-value from evaluate network for latent representation $\chi$, position $p$ and action $a$ at time $t$, and let $Q^*(\chi, p, a)$ denote the optimal action-value function, which we have calculated by Algorithm~\ref{alg:q_table}. The loss function could be described as follows:
\begin{equation}
\label{equation:loss_function}
L(\theta_i) =  L_{td} +\alpha KL(Q_t(\chi,p, \cdot; \theta_i)||Q^*(\chi,p, \cdot)) 
\end{equation}
where 
\begin{equation}
 L_{td} = \left( r + \gamma \max Q_t(\chi',a, \cdot; \theta_i') - Q_t(\chi,p,a;\theta_i) \right)^2
\end{equation} representing the TD error in DDQN and $\alpha$ is a coefficient that decays along time.
\begin{algorithm}[tb]
\caption{Construction of Optimal Action Value}
\label{alg:q_table}
\textbf{Input}: Multivariate Time Series $\mathcal{D}$ with Length $N$, Commission Fee Rate $\delta$, Action Space $A$ \\
\textbf{Output}: A Table $Q^*$ Indicating Optimal Action Value at Time $t$, Position $p$ and Action $a$. 
\begin{algorithmic}[1] 
\STATE Initialize $Q^*$ with shape (N,$|A|$,$|A|$) and all elements 0.
\FOR{$t \gets N-1$ to 1}  
\FOR{$p \gets 1$ to $|A|$}
\FOR{$a \gets 1$ to $|A|$}
\STATE  $Q^*[t,p,a]\gets\max_{a'} Q^*[t+1,a,a']+a\times p_{t+1}^{b1}-(p\times p_{t}^{b1}+E_{t}(p-a))$.
\ENDFOR
\ENDFOR
\ENDFOR
\STATE \textbf{return} $Q^*$
\end{algorithmic}
\end{algorithm}
The second term in Equation~\ref{equation:loss_function} enables low-level agents to acquire the advantage function of other actions under the same state without exploration, enhancing the efficiency of RL training. It can be proven that with this supervisor, the action value still converges to the optimal action value, as shown in Appendix C.2. 

\noindent
\textbf{Optimal Actor.} 
Although we have improved the training efficiency using the optimal value supervisor, it is still very hard for the agent to learn the optimal policy. The reason is that our optimal action value is based on the current position. It is often the case that once our RL agents deviate from the optimal policy, the supervision term also changes, which leads to a more significant deviation. Therefore, instead of just training from the transitions the agent explores using $\epsilon$-greedy policy, we further train the agent using the transitions generated by the optimal policy where the action with the highest optimal action value is chosen. The optimal transitions provide extra experience and prevent the agents from falling into the local trap. 

In Algorithm~\ref{alg:Efficient_RL}, the agents first collect experience from both $\epsilon$-greedy policy and the optimal actor, then update the network using both TD errors and KL divergence. 
\begin{algorithm}[tb]
\caption{Efficient RL with Q-Teacher}
\label{alg:Efficient_RL}
\textbf{Input}: Multivariate Time Series $\mathcal{D}$ with Length $N$, Commission Fee Rate $\delta$, Action Space $A$

\textbf{Output}: Network Parameter $\theta$

\begin{algorithmic}[1] 
\STATE Initialize experience replay $R$, network $Q_\theta$, target network $Q_{\theta'}$ and construct the optimal action value using Algorithm~\ref{alg:q_table} and trading environment $Env$.
\STATE Initialize trading environment $Env$
\FOR{$t = 1$ to $N-1$}
        \STATE Choose action $a_\epsilon$ using $\epsilon$-greedy policy.
        \STATE Store transition $(s, a_\epsilon, r, s',Q^*)$ in $D$
\ENDFOR
\STATE Reinitialize trading environment $Env$
\FOR{$t = 1$ to $N-1$}
        \STATE Choose action $a_o$ that $\argmax_{a} Q^*[t,p,a]$.
        \STATE Store transition $(s, a_o, r, s',Q^*)$ in $R$
\ENDFOR
        \STATE Sample transitions $(s_j, a_j, r_j, s'_j,Q^{*}_j)$ 
        \STATE Calculate $L$ following Equation~\ref{equation:loss_function}, do its gradient descent on $\theta$ and update $\theta'=\tau \theta+(1-\tau)\theta'$.
        
\STATE \textbf{return} $Q_{\theta}$
\end{algorithmic}
\end{algorithm}

\subsection{Stage II: Construction of Agent Pool}
The micro-level market information in the Crypto market changes rapidly, causing models' failure in maintaining their performance over a long period. According to our preliminary experiments where training among different market trends is incompatible, we decide to decompose the whole market as different trends and develop a suitable trading strategy for each market trend. 
\begin{algorithm}[th]
\caption{Market Segmentation \& Labelling}
\label{alg:segement_label_market}
\textbf{Input}: A Time Series $\mathcal{D}$ with Length $N$\\
\textbf{Parameter}: Risk threshold $\theta$, Label number $M$\\
\textbf{Output}: Labels indicating the trend they belong to for every point in time series D
\begin{algorithmic}[1] 
\STATE $D' \gets \text{denoising high frequency noise } D$.
\STATE Divide $D'$ according to its extrema into segments $S$. 
\STATE Merge adjacent segments in $S$ if DTW~\cite{muda2010voice} and slop difference are small enough until $S$ is stable.

\STATE Calculate threshold $H= Q_{1-\frac{\theta}{2}}(R)$, $L= Q_{\frac{\theta}{2}}(R)$ \;
\STATE Calculate the upper bond and lower bonds of slopes for each label based on the quantile and the threshold.
\STATE Label each segment based on the bonds.
\STATE Return the label corresponding to each segment.
\end{algorithmic}
\end{algorithm}

\noindent
\textbf{Generating Diverse Agents.} 
Previous works on generating diverse agents mainly focus on the different random seed initialization of the neural network or RL training's hyperparameter search~\cite{sun2023mastering}, which is mainly unstructured and can be seen as a byproduct of algorithmic stochasticity rather than an intentional design. Here we propose to train diverse agents following Algorithm~\ref{alg:Efficient_RL} with different preferences over time series $D$, i.e., market trends. We first separate the training dataset (a multivariate time series with a length of over 3 million) into data chunks with length $L$, where each data chunk represents a continuous market trend, to reduce the time horizon for training. The preference is defined by $\beta$, where a corresponding priority proportional to the probability of a data chunk with buy and hold return rate $r$ being sampled is calculated as Equation~\ref{equation:PES_definition}. 
\begin{equation}
f(x) = 
\begin{cases}
\frac{e^{\beta r}}{pdf(r)} & \text{if } Q_{\frac{\theta}{2}}(R) \leq r \leq Q_{1-\frac{\theta}{2}}(R) \\
e^{\beta r} & \text{if } r\geq Q_{1-\frac{\theta}{2}}(R) 
\lor r\leq Q_{\frac{\theta}{2}}(R)
\end{cases}
\label{equation:PES_definition}
\end{equation}
In Equation~\ref{equation:PES_definition}, \(Q_{\frac{\theta}{2}}(R)\) represents the \(\frac{\theta}{2}\)-th quantile of the samples' return rate \(R\). \(pdf\) represents probability density functions estimated by kernel density estimation and are calculated as Equation~\ref{eq:kernel}:
\begin{equation}
\label{eq:kernel}
    pdf(x)=  \frac{1}{nh} \sum_{r \in R} K\left(\frac{x - r}{h}\right)
\end{equation}
where $h$ is obtained by searching around Silverman's bandwidth~\cite{silverman1984spline} and $K$ is the kernel function as which we use the normal distribution \(N(0,1)\). The kernel density term erases the influence of the distribution of the training dataset and therefore provides a more robust sampling outcome.  We sample the data chunk based on the priority to construct our low-level MDP and train the agents following the process in stage I. Different agents are trained under different preference parameters $\beta$. This sampling method ensures the agent can access all of the data chunks yet is trained with a preference over all the market trends and prevents the agent from being trapped in those extreme conditions, which may cause agents' performances on all other trends to plummet.
\noindent
\textbf{Agent Selection.}
Although we have generated diverse agents, it is inefficient to put all of them into the agent pool because it will vastly increase the action space for the router. Therefore we only select a small fraction of generated agents to form the pool based on their profitability on various market trends. First, we precisely label each point in the valid dataset using Algorithm~\ref{alg:segement_label_market}, which, unlike previous algorithms~\cite{purkayastha2012diversification}, can label different datasets without tunning the hyperparameters. A more detailed version of the algorithm is described in Appendix C.1. We evaluate agents with different market trends and initial positions and further select the agents with the best profitability (averaged return on various market segments) under each label with each initial position to construct a two-dimensional agent pool $(m,n)$, where $m$ is the number of market trends and $n$ is the initial position.

\subsection{Stage III: Dynamic Routing Optimization}

We apply DDQN~\cite{van2016deep} to train the router for the high-level MDP. However, the number of agents in the pool is still too large. Even though the trajectory length has been significantly reduced (by 98.33\% \footnote{$1-\frac{59}{60}=0.9833$}) because we use the router to select the agent in a minute-level timescale, it is still computational-burdensome for the high-level agent to explore all the low-level agents. Therefore we use the priory knowledge of the agent pool to refine our options during trading. More specifically, before we choose the low-level agent, we will secure the chosen model whose initial positions are the same as the current position. Therefore, we reduce the number of possible low-level agents to $m$. Here, we choose not to compute a Q-teacher to aid the learning process for two reasons: i) the time horizon is largely reduced, therefore the computational burden for RL to self-explore is reduced. ii) the high-level action is a low-level agent, i.e., a trading strategy instead of a target position, causing the extra computation for the position at the end of the trading session of the selected agent and the reward during the trading session. The decreasing computation for RL and increasing computation for computing the optimal action value make pure DDQN more efficient.
\section{Experiment Setup}

\subsection{Datasets}
To comprehensively evaluate the algorithm, testing is conducted on four Crypto, encompassing both mainstream and niche options, over a period exceeding a week, covering both bull and bear market conditions. We summarize statistics of the 4 datasets in Table {\ref{tab:dataset}} and further elaborate them in Appendix D.1. For dataset split, we use data from the last 9 days for testing, the penultimate 9 days for validation and the remaining for training on all 4 datasets. We first train multiple low-level agents on the training dataset, and segment and label the valid dataset for model selection. We further train the router on the training dataset again and evaluate it on the whole valid dataset to pick the best router, which will be tested in the testing dataset. Experimental results in Table ~\ref{tab:performance_tex} show the great performance of that EarnHFT under different market statuses despite the difference between the valid dataset and the test dataset as shown in Appendix D.2.
\begin{table}[!h]
\setlength\tabcolsep{2.8pt}
    \small
  \centering
  \renewcommand{\arraystretch}{1.2}

    \begin{tabular}{|l|cccc|}
    \hline
    \textbf{Dataset} & \textbf{Dynamics}  & \textbf{Seconds} & \textbf{From} & \textbf{To} \\ \hline
    BTC/TUSD & Sideways  & 4057140  & 23/03/30 & 23/05/15  \\
    BTC/USDT & Sideways  & 3884400  & 22/09/01 & 22/10/15  \\
    ETH/USDT & Bear  & 3970800  & 22/05/01 & 22/06/15  \\
    GALA/USDT & Bull  & 3970740  & 22/07/01 & 22/08/15  \\ 
    \hline
    \end{tabular}
    \caption{Dataset statistics detailing market, data frequency, number of stocks, trading days and chronological period\footnote{a}.}
    \label{tab:dataset}
\end{table}
\footnotetext{The dates are in the formats YY/MM/DD, where 23/03/30 indicates Mar 30th of year 2023.}

\begin{table*}[!th]
\centering
\renewcommand{\arraystretch}{1.2}
  \resizebox{0.95\textwidth}{!}{
\begin{tabular}{|c|c|c|ccc|cc|c|c|c|ccc|cc|}
\hline
\multicolumn{2}{ |c | }{} & \multicolumn{1}{ c | }{Profit} & \multicolumn{3}{ c | }{Risk-Adjusted Profit} & \multicolumn{2}{ c| }{Risk Metrics}    &\multicolumn{2}{ |c | }{} & \multicolumn{1}{ c | }{Profit} & \multicolumn{3}{ c | }{Risk-Adjusted Profit} & \multicolumn{2}{ c| }{Risk Metrics}    
\\ 
\cline{1-16} 
Market   & Model & TR(\%)$\uparrow$ & ASR$\uparrow$ & ACR$\uparrow$ & ASoR$\uparrow$ & AVOL(\%)$\downarrow$ & MDD(\%)$\downarrow$ &Market   & Model & TR(\%)$\uparrow$ & ASR$\uparrow$ & ACR$\uparrow$ & ASoR$\uparrow$ & AVOL(\%)$\downarrow$ & MDD(\%)$\downarrow$ \\
\hline
{\multirow{7}{*}{\rotatebox[origin=c]{0}{BTCU}}} & DRA & -4.56 & -4.28   & -19.57 & -4.65 &   42.25  & 9.24 &{\multirow{7}{*}{\rotatebox[origin=c]{0}{BTCT}}} & DRA & -2.65 & \textcolor[RGB]{0,128,255}{-4.82}    & \textcolor[RGB]{0,128,255}{-17.48} & -4.77 &   21.18 & 5.84 \\
&  PPO  & -3.61 & -5.25   & -22.74 & -5.71 &   27.76  & \textcolor[RGB]{0,128,255}{6.41}& &  PPO  &  -0.60  & -14.74 & -35.80 & \textcolor[RGB]{0,128,255}{-0.14} & \textcolor[RGB]{0,153,76}{1.59} &  \textcolor[RGB]{0,153,76}{0.65}\\
\cline{2-8} \cline{10-16}

& CDQNRP & \textcolor[RGB]{0,153,76}{-2.83} & \textcolor[RGB]{0,153,76}{-2.91} &\textcolor[RGB]{0,128,255}{-14.85}& \textcolor[RGB]{0,128,255}{-3.31} & 37.61 &  7.38& &
CDQNRP &\textcolor[RGB]{0,128,255}{-0.60} & -19.52 & -37.88& -0.74 & \textcolor[RGB]{255,0,185}{1.20}  & \textcolor[RGB]{255,0,185}{0.61}  \\ 
& DQN & -3.48 & -12.37 & -35.01 & -11.86 & \textcolor[RGB]{255,0,185}{11.57} &   \textcolor[RGB]{0,153,76}{4.09} && DQN & \textcolor[RGB]{0,153,76}{0.47} & \textcolor[RGB]{255,0,185}{4.21} & \textcolor[RGB]{255,0,185}{27.94} & \textcolor[RGB]{255,0,185}{1.14} & \textcolor[RGB]{0,128,255}{4.38}  & \textcolor[RGB]{0,128,255}{0.66}\\ 
\cline{2-8} \cline{10-16}

& MACD &   -6.07 & -10.11 & -25.17 & -8.05 & \textcolor[RGB]{0,153,76}{24.84} &   9.98 && MACD &  -4.02 & -5.80 &  -24.21 & -4.32 & 26.87  & 6.44 \\ 
& IV &\textcolor[RGB]{0,128,255}{-2.99}  & \textcolor[RGB]{0,128,255}{-3.78} & \textcolor[RGB]{0,153,76}{-14.24} & \textcolor[RGB]{0,153,76}{-3.27} & 31.35  & 8.32 && IV & -12.01 & -17.83 & -38.99 & -13.90 & 27.68 &  12.66\\ 
\cline{2-8}\cline{10-16}
& EarnHFT & \textcolor[RGB]{255,0,185}{0.72} & \textcolor[RGB]{255,0,185}{1.22} & \textcolor[RGB]{255,0,185}{10.77} & \textcolor[RGB]{255,0,185}{0.93}& \textcolor[RGB]{0,128,255}{27.08}  &\textcolor[RGB]{255,0,185}{3.07}  && EarnHFT & \textcolor[RGB]{255,0,185}{0.99} & \textcolor[RGB]{0,153,76}{1.34} & \textcolor[RGB]{0,153,76}{7.76} &\textcolor[RGB]{0,153,76}{1.01}&32.40  &5.61 \\
\hline

{\multirow{7}{*}{\rotatebox[origin=c]{0}{ETH}}} &  DRA & -33.37 & \textcolor[RGB]{0,128,255}{-9.06}   & -32.23 & -9.20 &   163.25 & 45.88 &{\multirow{7}{*}{\rotatebox[origin=c]{0}{GALA}}} & DRA & 10.56 &4.77   & 41.63 & 4.44 &   92.43  & 10.60\\
&  PPO  & -22.61 & -10.11   & \textcolor[RGB]{0,128,255}{-31.17} &  -10.39 &   96.12  & 31.17 && PPO  & \textcolor[RGB]{0,128,255}{10.56} &\textcolor[RGB]{0,128,255}{4.77}   & \textcolor[RGB]{0,128,255}{41.63} & \textcolor[RGB]{0,128,255}{4.44} &   92.43  & 10.60 \\
\cline{2-8}\cline{10-16}
 & CDQNRP & \textcolor[RGB]{0,128,255}{-6.82} & -24.41 & -40.19 & \textcolor[RGB]{0,128,255}{-3.11} & \textcolor[RGB]{255,0,185}{11.46}  & \textcolor[RGB]{255,0,185}{6.96}& & CDQNRP & 5.22 & 4.51 & 39.42 & 4.16 & \textcolor[RGB]{0,153,76}{47.27} &  \textcolor[RGB]{0,153,76}{5.41} \\ 
& DQN & -11.02 & -9.47 & -32.81 & -8.43 & \textcolor[RGB]{0,153,76}{47.76}  & \textcolor[RGB]{0,153,76}{13.79}  && DQN & 2.94 & 3.55 & 32.02 & 2.66 & \textcolor[RGB]{255,0,185}{34.08} & \textcolor[RGB]{255,0,185}{3.78}\\ \cline{2-8}\cline{10-16}

& MACD & \textcolor[RGB]{0,153,76}{-4.29} & \textcolor[RGB]{0,153,76}{-1.78} & \textcolor[RGB]{0,153,76}{-8.71} & \textcolor[RGB]{0,153,76}{-1.19} &  79.64  & 16.35& & MACD &  2.37 & 1.79 & 11.45 &  1.22 &   \textcolor[RGB]{0,128,255}{62.89} &    9.84 \\ 
& IV & -27.42 & -12.27 & -36.01 &  -9.00 & 99.67  & 33.96 && IV & \textcolor[RGB]{0,153,76}{13.95} & \textcolor[RGB]{0,153,76}{6.74} & \textcolor[RGB]{0,153,76}{55.67} &  \textcolor[RGB]{0,153,76}{5.41} & 81.79  & 9.91\\ \cline{2-8}\cline{10-16}
 & EarnHFT & \textcolor[RGB]{255,0,185}{4.52} &\textcolor[RGB]{255,0,185}{2.92}  &\textcolor[RGB]{255,0,185}{14.30}  &\textcolor[RGB]{255,0,185}{1.78}& \textcolor[RGB]{0,128,255}{67.92}  &\textcolor[RGB]{0,128,255}{13.89}& & EarnHFT & \textcolor[RGB]{255,0,185}{19.41} &\textcolor[RGB]{255,0,185}{9.77}  & \textcolor[RGB]{255,0,185}{79.08}  &\textcolor[RGB]{255,0,185}{7.79}&  74.94  & \textcolor[RGB]{0,128,255}{9.26}  \\ 
\hline
\end{tabular}
}
\caption{Performance comparison on 4 Crypto markets with 6 baselines including 2 policy-based RL algorithms, 2 value-based RL algorithms, and 2 rule-based methods. Results in pink, green, and blue show the best, second-best, and third-best results.}
\label{tab:performance_tex}
\end{table*}

\subsection{Evaluation Metrics}
We evaluate EarnHFT on 6 different financial metrics including one profit criterion, two risk criteria, and three risk-adjusted profit criteria listed below.
\begin{itemize}
    \item \textbf{Total Return (TR)} is the overall return rate of the whole trading period.
    It is defined as \( TR = \frac{V_{t}-V_{1}}{V_{1}} \), where \({V}_{t}\) is the final net value and \({V}_{1}\) is the initial net value.
    \item \textbf{Annual Volatility (AVOL)} is the variance of the annual return defined as \(\sigma[\mathbf{ret}] \times \sqrt{m}\) to measure the risk level of trading strategies, where \(\mathbf{ret}=(ret_1, ret_2,...,ret_t)\) is a vector of secondly return, $\sigma[.]$ is the variances and $m$ is the number of seconds contained in a year.
    \item \textbf{Drawdown (MDD)} measures the largest loss from any peak to show the worst case.
    \item \textbf{Annual Sharpe Ratio (ASR)} considers the amount of extra return that a trader receives per unit of increase in risk. It is defined axs: \( SR = {E}[\mathbf{ret} ]/\sigma[\mathbf{ret}] \times \sqrt{m}\), where \(E[.]\) is the expected value.
    \item \textbf{Annual Calmar Ratio (ACR)} is defined as \(CR = \frac{E[\mathbf{ret}]}{MDD} \times m \), which is the expected annual return divided by the maximum drawdown.
    \item \textbf{Annual Sortino Ratio (ASoR)} applies downside deviation as the risk measure. It is defined as: \(SoR = \frac{E[\mathbf{ret}]\times \sqrt{m}}{DD}\), where downside deviation is the standard deviation of the negative return rates.
\end{itemize}

\subsection{Training Setup}
We conduct all experiments on a 4090 GPU. For the trading setting, the commission fee rate is 0 for BTCT and 0.02\% for the remaining datasets following the policy of Binance. For the training setting, we choose $\beta$ in Equaition~\ref{equation:PES_definition} in list $[-90,-10,30,100]$ and run each $\beta$ for 50 epochs, generating a total of 200 agents. Adam is used as the optimizer for DDQN. As for other baselines, there are two conditions: i) there are authors’ official or open-source library \cite{huang2022cleanrl} implementations, we apply the same hyperparameters for a fair comparison\footnote{PPO and DQN.}. ii) if there are no publicly available implementations\footnote{DRA and CDQNRP}, we reimplement the algorithms and try our best to maintain consistency based on the original papers. It takes about 10 hours to run all experiments in 4 datasets. Descriptions of other parameter settings (e.g., the trading setting) are in Appendix D.3.

\subsection{Baselines}
To provide a comprehensive comparison of EarnHFT, we select 6 baselines including 4 SOTA RL algorithms and 2 widely-used rule base methods.

\begin{itemize}
    \item \textbf{PPO~\cite{schulman2017proximal}} applies importance sampling to enhance the experience efficiency.
    \item \textbf{DRA~\cite{briola2021deep}} uses an LSTM~\cite{hochreiter1997long} network to enhance the state representation to gain a better result using PPO.
    \item \textbf{DQN~\cite{mnih2015human}} applies experience replay and multi-layer perceptrons to Q-learning.
    \item \textbf{CDQNRP~\cite{zhu2022quantitative}} uses a random perturbed target frequency to enhance the stability during training.
    \item \textbf{MACD~\cite{krug2022enforcing}} is an upgraded method based on the traditional moving average method. Not only does it show the rise or fall of the current price, but also indicates the speed of rising or falling.
    \item \textbf{IV~\cite{chordia2002order}} is a micro-market indicator widely used in HFT.
\end{itemize}

\section{Results and Analysis}
\subsection{Comparison with Baselines}
\begin{figure}[!th]
\begin{center}
\includegraphics[width=0.45\textwidth]{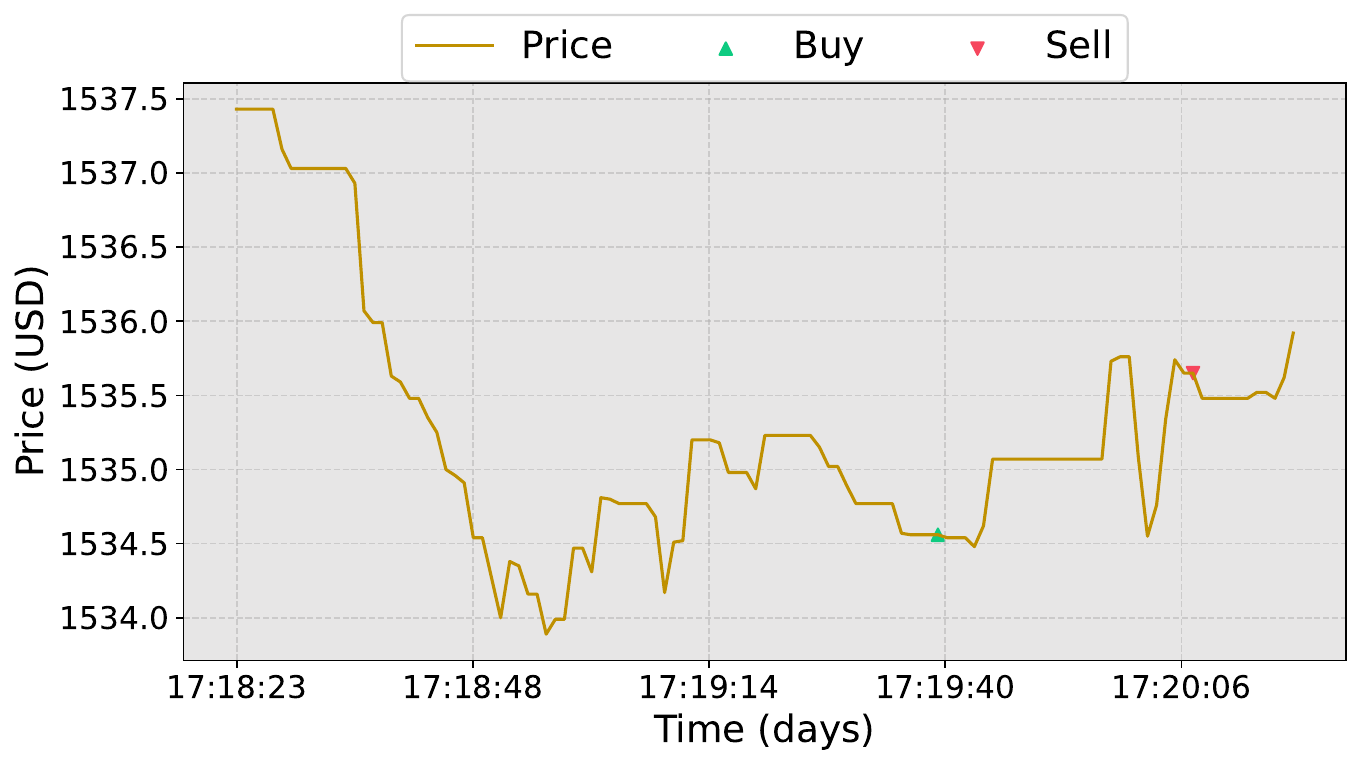}
\end{center}
\caption{Trading process of EarnHFT in ETH}
\label{fig:Trading_Process}
\end{figure}

According to Table~\ref{tab:performance_tex}, our method achieves the highest profit in all 4 datasets and the highest risk-adjusted profit in 3 datasets. Value-based methods (e.g., CDQRP and DQN) perform well when the gap between the valid and test datasets is not large under a stable market trend.
Policy-based methods (e.g., PPO and DRA) are easy to converge to a dummy policy where the agents just deliver the target position the same as their current position due to the existence of the commission fee even if the learning rate is set to $1e^{-7}$ and therefore perform poorly on the bear market. Rule-based methods are extremely sensitive to the take profit point and the stop loss and only achieve moderate profit under volatile markets. Our method, EarnHFT, although it performs well on profit-related metrics, is a very radical trader due to the optimal value supervisor and optimal actor, which only delivers profit-related experience, neglecting the risk-related information, and therefore performs moderately in some datasets in terms of risk. As shown in Figure~\ref{fig:Trading_Process}, EarnHFT opens a position and closes the position within 30 seconds and profits from a market trend which is viewed as a pullback at a minute-level timescale. More results can be found in Appendix D.4.

\subsection{The Effectiveness of Hierarchical Framework}
\begin{figure}[!th]
\begin{center}
\includegraphics[width=0.48\textwidth]{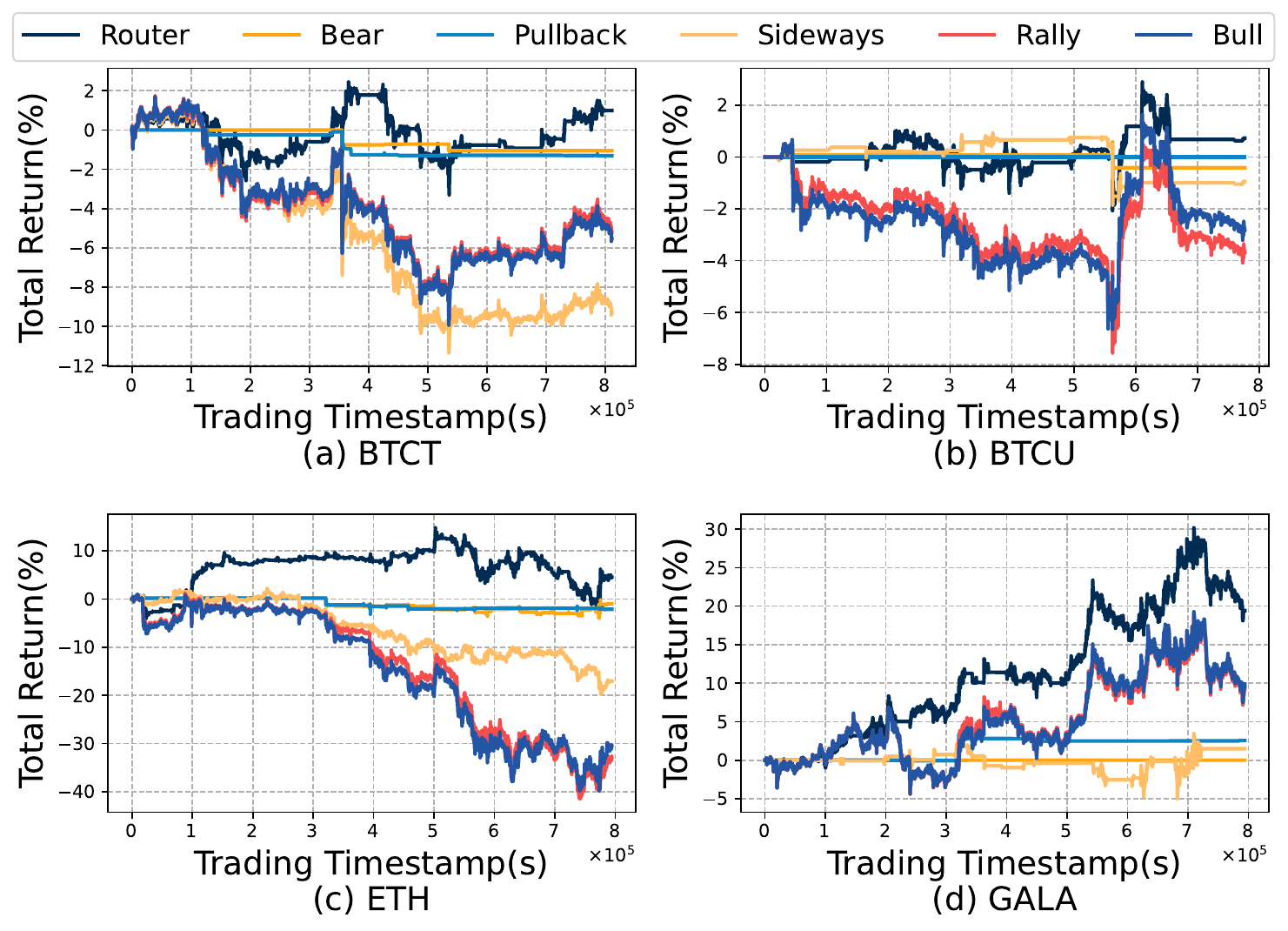}
\end{center}
\caption{Comparison of the router and agent pool}
\label{fig:Ablation_of_4_dataset}
\end{figure}

\begin{figure}[!th]
\begin{center}
\includegraphics[width=0.41\textwidth]{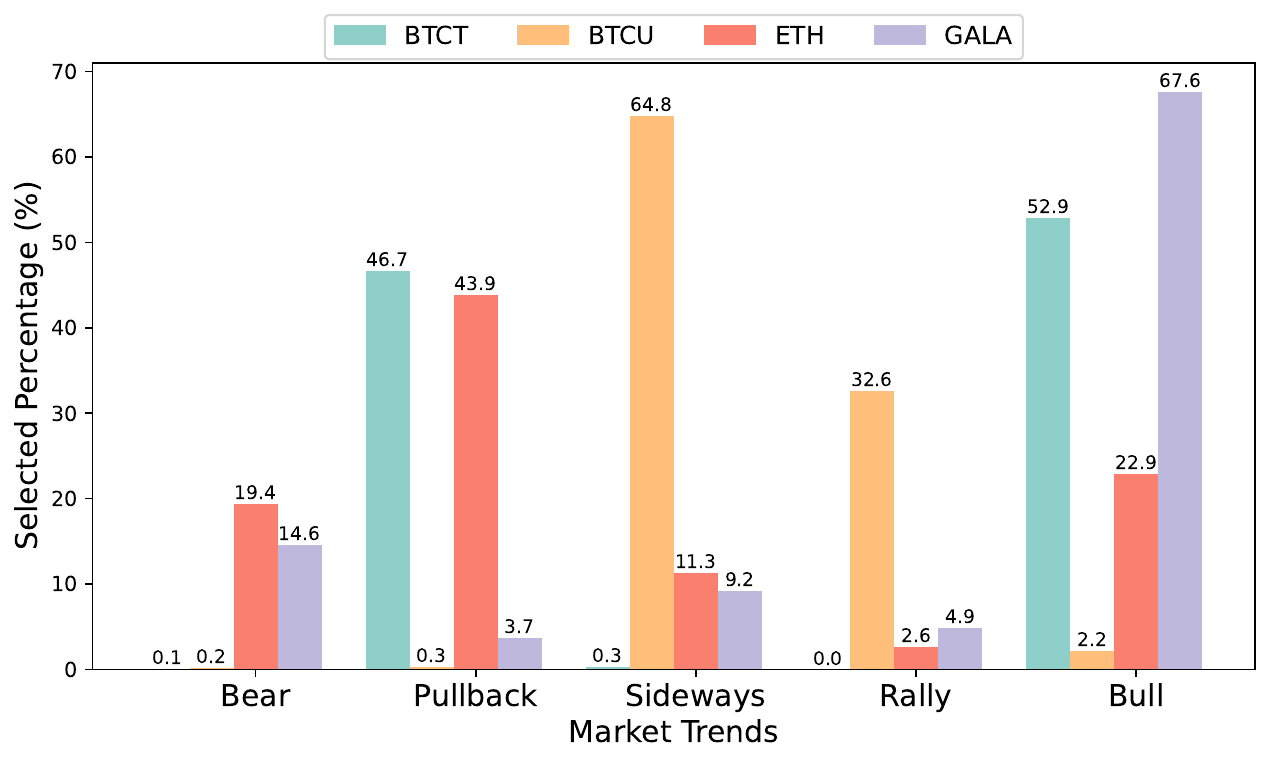}
\end{center}
\caption{Router selection distribution}
\label{fig:Selection_of_4_dataset}
\end{figure}

We examine the effectiveness of the hierarchical framework by analyzing the router's behaviors under different datasets and conducting experiments to show the performance comparison of the EarHFT and each agent from its pool. From Figure~\ref{fig:Ablation_of_4_dataset} we can see that bull and rally agent tends to buy and hold and therefore perform well in the bull market (e.g., GALA). The sideways agent tends to trade less and hold still its position. The pullback and bear market tends to close its position and perform well in the bear market (e.g., ETH). The router combines all the advantages of the agents and performs the best on all 4 datasets in terms of profit. Figure~\ref{fig:Selection_of_4_dataset} refers to the selection distribution for the router on 4 datasets. While datasets with high volatility (e.g., ETH and GALA), the market dynamics change more frequently and therefore the routing shows a more balanced distribution across 5 market trends. While datasets with lower volatility (e.g., BTCT and BTCU), the router selection is more focused on two market dynamics.

\subsection{The Effectiveness of Optimal Action Value}
\begin{table}[htb]
\setlength\tabcolsep{4pt}
  \begin{center}
    \begin{tabular}{|cc|ccc|ccc|}
    \hline
      \multirow{2}{*}{OA} &\multirow{2}{*}{OS} & \multicolumn{3}{c|}{GALA} & \multicolumn{3}{c|}{ETH}\\
      & & CS & RS& AHL & CS & RS & AHL \\
      \hline
      \checkmark&\checkmark  & 78848  &4.43 &448 & 102400 & 12.32 & 81.3 \\
       \checkmark&  &   102400 & 0.24 &38.7 & 102400 & -1.40 & 4.15 \\
        &  \checkmark&   4608&2.89 &147 & 30720  & 4.87 & 35.8  \\
         &  &   30720& -0.01& 284& 30720 & -29.6  & 39.1  \\
         \hline
    \end{tabular}
  \end{center}
  \caption{Ablation Study of OS and OT}
  \label{table:ablation_os_ot}
\end{table}
To demonstrate the effectiveness of the optimal value supervisor (OS) and the optimal actor (OA), we conduct an ablation study on two datasets, ETH and GALA. We evaluate the training efficiency by the number of steps need to converge (CS) and the converged reward sum (RS). We further investigate their influence on agent trading behavior by average holding length (AHL). In Table~\ref{table:ablation_os_ot}. For GALA we can see that compared with the original DDQN, the one with OS only takes 15\% of the steps to converge and gain a higher return. OA can further improve the return in exchange for more steps to converge. For ETH, since the market is bull, the CS is not reduced by OS. However, the return is largely increased. The reason why OS is more effective is that OS provides more information for the agents and its instructions vary along the changes of the agents' policy while OA only provides demonstrations.

\section{Conclusion}
In this paper, we propose EarnHFT, a novel three-stage hierarchical RL framework for HFT to alleviate training efficiency and data shifting. First, we compute the optimal action value to improve the performance and training efficiency of second-level RL agents. Then we train a diverse pool of agents excelling in various market trends. Finally, we train a router to regularly pick an agent from the pool to conduct trading to deal with the dynamic market. Extensive experiments on Crypto markets demonstrate that EarnHFT significantly outperforms many strong baselines. Ablation studies show the effectiveness of the proposed components.
\clearpage


\bibliography{fin}

\begin{thebibliography}{28}
\providecommand{\natexlab}[1]{#1}

\bibitem[{Almeida and Gonçalves(2023)}]{ALMEIDA2023100785}
Almeida, J.; and Gonçalves, T.~C. 2023.
\newblock A systematic literature review of investor behavior in the
  cryptocurrency markets.
\newblock \emph{Journal of Behavioral and Experimental Finance}, 37: 100785.

\bibitem[{Briola et~al.(2021)Briola, Turiel, Marcaccioli, and
  Aste}]{briola2021deep}
Briola, A.; Turiel, J.; Marcaccioli, R.; and Aste, T. 2021.
\newblock Deep reinforcement learning for active high frequency trading.
\newblock \emph{arXiv preprint arXiv:2101.07107}.

\bibitem[{Chordia, Roll, and Subrahmanyam(2002)}]{chordia2002order}
Chordia, T.; Roll, R.; and Subrahmanyam, A. 2002.
\newblock Order imbalance, liquidity, and market returns.
\newblock \emph{Journal of Financial Economics}, 65(1): 111--130.

\bibitem[{Cumming, Alrajeh, and Dickens(2015)}]{cumming2015investigation}
Cumming, J.; Alrajeh, D.~D.; and Dickens, L. 2015.
\newblock An investigation into the use of reinforcement learning techniques
  within the algorithmic trading domain.
\newblock \emph{Imperial College London: London, UK}, 58.

\bibitem[{Delfabbro, King, and Williams(2021)}]{delfabbro2021psychology}
Delfabbro, P.; King, D.~L.; and Williams, J. 2021.
\newblock The psychology of cryptocurrency trading: Risk and protective
  factors.
\newblock \emph{Journal of Behavioral Addictions}, 10(2): 201--207.

\bibitem[{Hochreiter and Schmidhuber(1997)}]{hochreiter1997long}
Hochreiter, S.; and Schmidhuber, J. 1997.
\newblock Long short-term memory.
\newblock \emph{Neural computation}, 9(8): 1735--1780.

\bibitem[{Huang, Huang, and Ni(2019)}]{huang2019predicting}
Huang, J.-Z.; Huang, W.; and Ni, J. 2019.
\newblock Predicting bitcoin returns using high-dimensional technical
  indicators.
\newblock \emph{The Journal of Finance and Data Science}, 5(3): 140--155.

\bibitem[{Huang et~al.(2022)Huang, Dossa, Ye, Braga, Chakraborty, Mehta, and
  Ara{\'u}jo}]{huang2022cleanrl}
Huang, S.; Dossa, R. F.~J.; Ye, C.; Braga, J.; Chakraborty, D.; Mehta, K.; and
  Ara{\'u}jo, J.~G. 2022.
\newblock Cleanrl: High-quality single-file implementations of deep
  reinforcement learning algorithms.
\newblock \emph{The Journal of Machine Learning Research}, 23(1): 12585--12602.

\bibitem[{Krug, Dobaj, and Macher(2022)}]{krug2022enforcing}
Krug, T.; Dobaj, J.; and Macher, G. 2022.
\newblock Enforcing network safety-margins in industrial process control using
  MACD indicators.
\newblock In \emph{European Conference on Software Process Improvement},
  401--413.

\bibitem[{Madhavan(2000)}]{madhavan2000market}
Madhavan, A. 2000.
\newblock Market microstructure: A survey.
\newblock \emph{Journal of Financial Markets}, 3(3): 205--258.

\bibitem[{Mnih et~al.(2013)Mnih, Kavukcuoglu, Silver, Graves, Antonoglou,
  Wierstra, and Riedmiller}]{mnih2013playing}
Mnih, V.; Kavukcuoglu, K.; Silver, D.; Graves, A.; Antonoglou, I.; Wierstra,
  D.; and Riedmiller, M. 2013.
\newblock Playing Atari with deep reinforcement learning.
\newblock \emph{arXiv preprint arXiv:1312.5602}.

\bibitem[{Mnih et~al.(2015)Mnih, Kavukcuoglu, Silver, Rusu, Veness, Bellemare,
  Graves, Riedmiller, Fidjeland, Ostrovski et~al.}]{mnih2015human}
Mnih, V.; Kavukcuoglu, K.; Silver, D.; Rusu, A.~A.; Veness, J.; Bellemare,
  M.~G.; Graves, A.; Riedmiller, M.; Fidjeland, A.~K.; Ostrovski, G.; et~al.
  2015.
\newblock Human-level control through deep reinforcement learning.
\newblock \emph{Nature}, 518(7540): 529--533.

\bibitem[{Muda, Begam, and Elamvazuthi(2010)}]{muda2010voice}
Muda, L.; Begam, M.; and Elamvazuthi, I. 2010.
\newblock Voice recognition algorithms using mel frequency cepstral coefficient
  (MFCC) and dynamic time warping (DTW) techniques.
\newblock \emph{arXiv preprint arXiv:1003.4083}.

\bibitem[{Niu, Li, and Li(2022)}]{niu2022metatrader}
Niu, H.; Li, S.; and Li, J. 2022.
\newblock MetaTrader: An reinforcement learning approach integrating diverse
  policies for portfolio optimization.
\newblock In \emph{Proceedings of the 31st ACM International Conference on
  Information \& Knowledge Management}, 1573--1583.

\bibitem[{Purkayastha, Manolova, and
  Edelman(2012)}]{purkayastha2012diversification}
Purkayastha, S.; Manolova, T.~S.; and Edelman, L.~F. 2012.
\newblock Diversification and performance in developed and emerging market
  contexts: A review of the literature.
\newblock \emph{International Journal of Management Reviews}, 14(1): 18--38.

\bibitem[{Schulman et~al.(2017)Schulman, Wolski, Dhariwal, Radford, and
  Klimov}]{schulman2017proximal}
Schulman, J.; Wolski, F.; Dhariwal, P.; Radford, A.; and Klimov, O. 2017.
\newblock Proximal policy optimization algorithms.
\newblock \emph{arXiv preprint arXiv:1707.06347}.

\bibitem[{Shiryaev(2008)}]{shiryaev2008thou}
Shiryaev, X.~Y. 2008.
\newblock Thou shalt buy and hold.
\newblock \emph{Quantitative Finance}, 8(8): 765--776.

\bibitem[{Silverman(1984)}]{silverman1984spline}
Silverman, B.~W. 1984.
\newblock Spline smoothing: the equivalent variable kernel method.
\newblock \emph{The Annals of Statistics}, 898--916.

\bibitem[{Sun et~al.(2023)Sun, Wang, Xue, Lou, and An}]{sun2023mastering}
Sun, S.; Wang, X.; Xue, W.; Lou, X.; and An, B. 2023.
\newblock Mastering stock markets with efficient mixture of diversified trading
  experts.
\newblock \emph{Proceedings of the 29th ACM SIGKDD International Conference on
  Knowledge Discovery \& Data Mining}.

\bibitem[{Sun et~al.(2022)Sun, Xue, Wang, He, Zhu, Li, and
  An}]{sun2022deepscalper}
Sun, S.; Xue, W.; Wang, R.; He, X.; Zhu, J.; Li, J.; and An, B. 2022.
\newblock DeepScalper: A Risk-aware reinforcement learning framework to capture
  fleeting intraday trading opportunities.
\newblock In \emph{Proceedings of the 31st ACM International Conference on
  Information \& Knowledge Management}, 1858--1867.

\bibitem[{Sutton and Barto(2018)}]{sutton2018reinforcement}
Sutton, R.~S.; and Barto, A.~G. 2018.
\newblock \emph{Reinforcement learning: An introduction}.
\newblock MIT press.

\bibitem[{Th{\'e}ate and Ernst(2021)}]{theate2021application}
Th{\'e}ate, T.; and Ernst, D. 2021.
\newblock An application of deep reinforcement learning to algorithmic trading.
\newblock \emph{Expert Systems with Applications}, 173: 114632.

\bibitem[{Van~Hasselt, Guez, and Silver(2016)}]{van2016deep}
Van~Hasselt, H.; Guez, A.; and Silver, D. 2016.
\newblock Deep reinforcement learning with double q-learning.
\newblock In \emph{Proceedings of the AAAI Conference on Artificial
  Intelligence}.

\bibitem[{Wang et~al.(2021)Wang, Wei, An, Feng, and Yao}]{wang2021commission}
Wang, R.; Wei, H.; An, B.; Feng, Z.; and Yao, J. 2021.
\newblock Commission fee is not enough: A hierarchical reinforced framework for
  portfolio management.
\newblock In \emph{Proceedings of the AAAI Conference on Artificial
  Intelligence}, volume~35, 626--633.

\bibitem[{Zhang et~al.(2023{\natexlab{a}})Zhang, Duan, Chen, Chen, Li, and
  Zhao}]{zhang2023towards}
Zhang, C.; Duan, Y.; Chen, X.; Chen, J.; Li, J.; and Zhao, L.
  2023{\natexlab{a}}.
\newblock Towards Generalizable Reinforcement Learning for Trade Execution.
\newblock \emph{arXiv preprint arXiv:2307.11685}.

\bibitem[{Zhang et~al.(2023{\natexlab{b}})Zhang, Duan, Chen, Chen, Li, and
  Zhao}]{zhang2023generalizable}
Zhang, C.; Duan, Y.; Chen, X.; Chen, J.; Li, J.; and Zhao, L.
  2023{\natexlab{b}}.
\newblock Towards generalizable reinforcement learning for trade execution.
\newblock arXiv:2307.11685.

\bibitem[{Zhou et~al.(2021)Zhou, Qin, Torres, Le, and Gervais}]{zhou2021high}
Zhou, L.; Qin, K.; Torres, C.~F.; Le, D.~V.; and Gervais, A. 2021.
\newblock High-frequency trading on decentralized on-chain exchanges.
\newblock In \emph{2021 IEEE Symposium on Security and Privacy (SP)}, 428--445.

\bibitem[{Zhu and Zhu(2022)}]{zhu2022quantitative}
Zhu, T.; and Zhu, W. 2022.
\newblock Quantitative trading through random perturbation Q-network with
  nonlinear transaction costs.
\newblock \emph{Stats}, 5(2): 546--560.

\end{thebibliography}
\end{document}


\appendix

\section{More detailed Related Work}
\label{sec:app_related_work}
\subsection{High-Frequency Trading in Crypto}
High-frequency trading (HFT) is an algorithmic trading that involves high speed trading execution within seconds\cite{cartea2015algorithmic}. This method is mainly adopted by institutional investors. The main idea of high frequency trading is to build a mathematical \& statistical model based on the micro-level of the market and let the computer conduct the order placement, cancellation or execution based on the result of the model. Compared with hand-crafted trading, algorithm trading could work 24/7 and execute order efficiently and motionlessly, avoiding mistakes that human could make.   

In the cryptocurrency market, there are many technical indicators\cite{huang2019predicting} to help make better decisions. Those technical indicators are derived from time series analysis or simply from market events.   
For example, Imbalance volume\cite{chordia2002order} is a volume-based technical indicator that can be used to measure buying and selling pressure of assets, which is calculated as 
\begin{equation}
    imblance=\frac{\sum_i(ask\_size_i-bid\_size_i) }{\sum_i(ask\_size_i+bid\_size_i)}
\end{equation} where $ask\_size_i$ indicates the $i^{th}$ level's size on the sell side. The value' range is [-1,1]. When value approaches to -1, it indicates that there are a lot of orders demanding to buy this and few orders demanding to sell, indicating the price will go up. When value approaches to 1, it indicates that there are a lot of orders demanding to sell and few orders demanding to buy, indicating the price will go down.
\cite{krug2022enforcing} proposes a finance indicator as a varaint of moving average called moving average convergence divergence (MACD) which is calculated as 
\begin{equation}
    MACD= DIF - DEA 
\end{equation}
\begin{equation}
    DIF=P.EMA(mid)-P.EMA(long)
\end{equation} 
\begin{equation}
    DEA=DIF.EMA(short)
\end{equation} 
where $P$ is the price and $X.EMA(L)$ indicates the exponential moving average of series X with window length L.
It captures trends among different time scales. When $MACD > 0$ and $DIF > 0$, it means that not only the price goes up, the speed of rising also goes up, so we should buy the asset at this moment. When $MACD \times DIF <= 0$, it is either the price goes up but the speed of rising also goes down or the price goes down but the speed of falling also goes down. In this moment, the situation is not clear and therefore we hold our position for further signal. When $MACD < 0$ and $DIF < 0$, mean that price goes down and speed of falling goes up. It means the price of the asset will continue decrease and therefore we sell.

\subsection{Reinforcement Learning for HFT}
\label{sec:app_related_work:RL for HFT}
In general, the related algorithms could be classified into  value-based algorithms and policy-based algorithms.

\noindent
\textbf{value-based algorithms} \cite{shin2019deep} applies LSTM and CNN to DQN to do PM on daily base.
 \cite{nagy2023asynchronous} applies D3QN with Apex to Lobster dataset with a modified reward function.\cite{takara4411793deep} applies distributional DQN with hindsight experience replay and pre-processing sampling to high frequency trading.\cite{zhu2022quantitative} applies convolution DQN with a random perturbation method to high frequency trading for stability.

\noindent
\textbf{policy-based algorithms} 
\cite{jiang2017cryptocurrency} proposes convolutional neural network reinforcement learning with deep policy gradient (DPG) algorithm to conduct PM on mintue level data. \cite{si2017multi} applies a long-short-term-memory (LSTM) network into deep deterministic policy gradient (DDPG) algorithm and modified the reward function to let the agent balance between risk and profit.\cite{conegundes2020beating} applies DDPG on daily level algorithm trading. 
\cite{briola2021deep} propose a PPO-based algorithms with lstm network. Transitions containing large price change are used to train the agent to gain a good result.\cite{liu2021bitcoin} use a risk-adjusted profit as reward to gain a conservative result and use Bayesian Optimization to discover hyper-parameter combinations.

\subsection{Hierarchical RL for Quantitative Trading}
\label{sec:app_related_work:Hierarchical RL for Quantitative Trading}
Hierarchical Reinforcement Learning (HRL), which decomposes a long-horizon task into a hierarchy of sub-problems, has been studied for decades. Its capability of handling situation from different level has drawn traders' attention. ~\cite{zha2022hierarchical} proposes a hierarchical reinforcement learning framework for better refining the stock pool. The high level agent selects several stocks with high profitable probability among a large stock pool while the low agent tries to optimize the portfolio weights among those selected stocks. \cite{wang2021commission} proposes a hierarchical reinforcement learning framework for better simulation of trading process. The high level agents optimize the portfolio weights among a fixed stock pool while the low level agents optimize the order execution process. 
\cite{gao2021framework} introduces a framework, based on the hierarchical Deep Q-Network, that addresses the issue of zero commission fee by reducing the number of assets assigned to each deep Q-Network and dividing the total portfolio value into smaller parts. Furthermore, this framework is
flexible enough to handle an arbitrary number of assets.





\section{Problem Formulation}
\label{sec:appendix_pf}
In this section, we will first present some basic finance concepts and the definition of the MDP. Then we will introduce the hierarchical MDP formulation in detail.

\subsection{Basic Financial Concepts}
\label{sec:finance_concept}
In this section, we will talk about the basic financial concepts in details which will be used in the hierarchical MDP formulation.

\begin{figure}[th]
\begin{center}
\vspace{-0.2cm}
\includegraphics[width=0.48\textwidth]{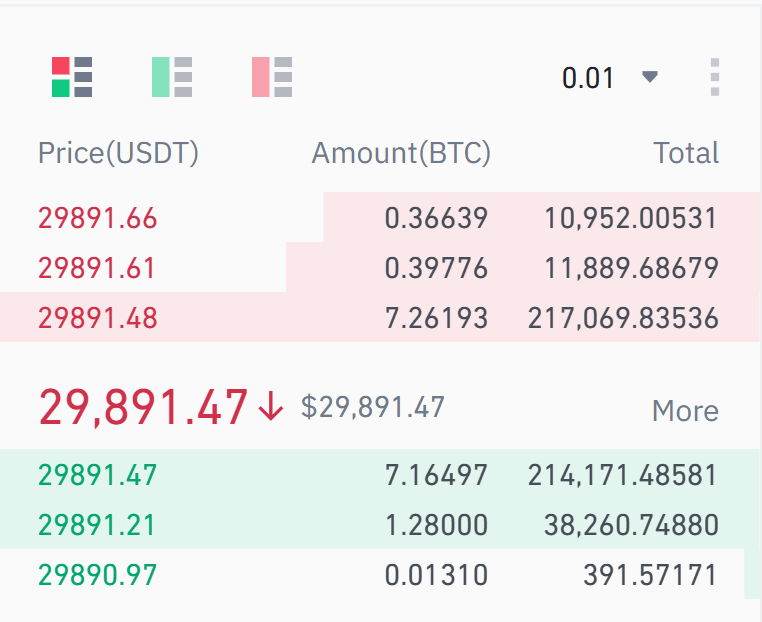}
\end{center}
\vspace{-0.2cm}
\caption{Snapshot of LOB}
\vspace{-0.2cm}
\label{fig:Snapshot_of_LOB}
\end{figure}

\noindent
\textbf{LOB} is an aggregated record of unfilled orders by price and side as shown in Figure \ref{fig:Snapshot_of_LOB}. In the upper part, it shows the selling side, which means the price that a seller want to sell and its corresponding amount. In the lower part, it show the buying side, which means the price that a buyer want to buy and its corresponding amount. It shows the best available buying and selling prices and liquidity.

\begin{figure}[th]
\begin{center}
\vspace{-0.2cm}
\includegraphics[width=0.48\textwidth]{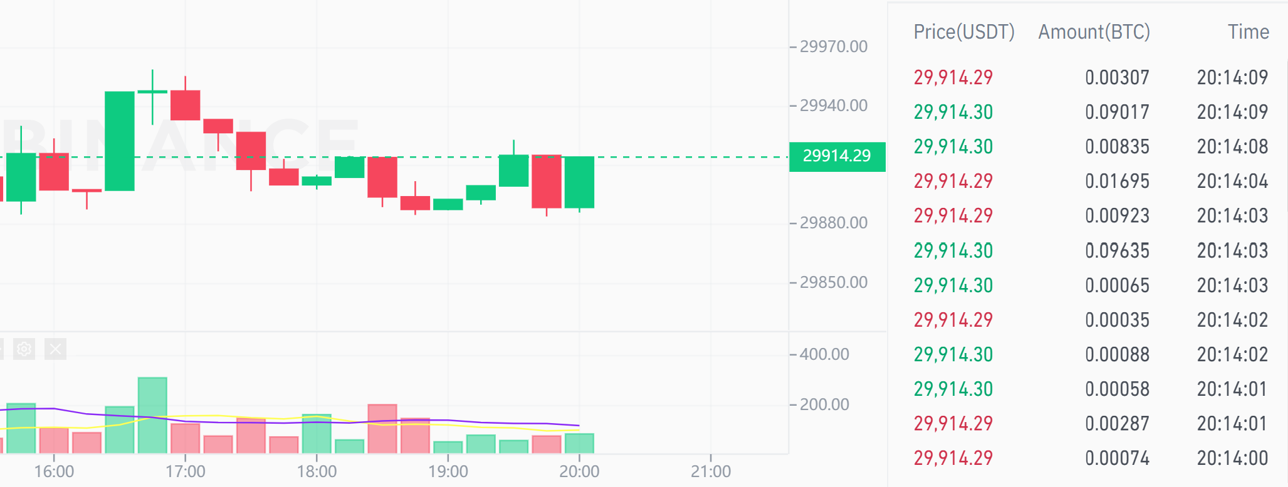}
\end{center}
\vspace{-0.2cm}
\caption{OHLC and executed order}
\vspace{-0.2cm}
\label{fig:Snapshot of OHLC_Trade}
\end{figure}

\noindent
\textbf{OHLC} is aggregated information of executed orders over a period. OHLC contains open, high low and close, indicating the first, highest, lowest and last price of all executed orders. As shown in Figure \ref{fig:Snapshot of OHLC_Trade}, on the left side picture is the candle chart, which is for describing the OHLC. The upper and lower end for the wicks are the high price and low price respectively. If the body is green, the upper and lower end for the bodies are the close price and open price respectively. If the body is red, the upper and lower end for the bodies are the open price and close price respectively. On the right side is the trade information, which includes the side executed order and the corresponding price and amount. 

\noindent
\textbf{Market Order} is an instruction to buy or sell a specific amount of financial instrument immediately at the best available current price, prioritizing execution speed over a specific price. It will take orders from the opposite side of the limit order book and execute the corresponding amount immediately. For example, in Figure \ref{fig:Snapshot of OHLC_Trade} the reason why the price of executed order is jumping between 29914.29 and 29914.30 is that side of the market order is jumping between selling and buying. If the side of the market order is buying, meaning that the trader wants to buy an amount of BTC immediately, then the broker will match its order to the opposite side of LOB, here more specifically is the selling side. The order size will be first compared with $ask\_1\_size$. If your order size is smaller than $ask\_1\_size$,  then all of your orders will be conducted with price $ask\_1\_price$, which means that your order is conducted at the lowest price from the selling side. However, it is non-trivial to point out that sometimes, especially when the price is changing rapidly, a single-level size might not be enough to fill the whole order. In that case, the rest of the order has to be conducted by other prices because the best price unfilled order has already been taken. That is why we introduce Equation \ref{equation:calculate_market_order}. It calculates the market order price level-by-level and therefore our simulated environment is more realistic than the trading environment in \cite{jiang2017deep}. It demonstrates a property: no matter whether you are buying or selling using a market order, the larger the size, the more likely you will get a less favorable price: Take LOB in Figure \ref{fig:Snapshot_of_LOB} as an example. if you are buying and your size does not exceed $ask\_1\_size$, then the unit conducted price is just $ask\_1\_price$, which is 29891.48. However, if your size exceeds $ask\_1\_size$ and yet does not exceed $ask\_1\_size+ask\_2\_size$, then you have to conduct the rest order using $ask\_2\_price$, whose unit price is 29891.61.  \cite{zhu2022quantitative} consider this problem as well, but since they only have trade information, it does not compute the trading cost precisely as we do. They construct a stochastic model, which depends on the order size and market condition, delivering a trading cost similar to the commission fee rate to alleviate the problem. Another thing worth mentioning is that even without the commission fee rate $\sigma$, the trading still has cost because of the ask-bid spread. Since you are using a market order, the price needed to buy an asset is slightly higher than the price needed to sell it, because you can only sell the asset to the people who want to buy it at a low price and buy the asset from the people who want to sell it with a high price. Although the spread of 0.01 does not seem much, yet it could cause tremendous losses if the trading speed is extremely high.
\begin{equation}
\label{equation:calculate_market_order}
E_t(M) = \sum_{i} (p_t^i \times \min(q_t^i, R_{i-1}))(1+\sigma)
\end{equation}
\subsection{Motivation for Hierarchical Formulation}
A key challenge to real-world high-frequency trading is to maintain its performance while the market dynamics change rapidly. In our preliminary experiments, extensive results have shown that one single agent fails to excel in all market conditions because training on different market dynamics is incompatible with each other. Ideal policies should be different for similar micro-level market conditions under different macro-level market conditions. In a bull market, facing a slight drop in the price does not mean we have to sell the asset right away, because the drop is only temporary and even later we bought it at a lower price, the profit does not cover the trading cost of selling it and buying it again because of the commission fee and ask bid spread, while in a bear market, and we buy something when there is bottoming out, we should sell it right away when the price goes down because the rise is only temporary and if you don't sell it right away, the price will not be as high and evenly we will lose the money. A direct concatenation of both macro-level and micro-level market information will not work for two reasons: 1) As our reward is defined as the second-level net value fluctuation, the correlation between the macro-level market information and the reward is not high, the macro-level market information will affect the performance of the micro-level agent. 2) We previously viewed the direct concatenation of both macro-level and micro-level market information as a goal-conditioned RL, and hope to train it in an HER way so that we can use less data to train a better agent because there are some common skills that a trader in different market could share. However, we found that the training efficiency is not improved because there are few skills that an ideal trader in a bull market and an ideal trader in a bear market could share and the unstable low-level agents makes the training of the high-level agent even harder. 
Therefore, we propose the hierarchical framework of high-frequency trading, aiming to solve the incompatible training among different market dynamics and integrate information with multi-time granularity for better trading. We first train a pool of fixed diverse agents and then use the pool to train a router that dynamically picks an agent suitable for the current market state. The fixed low-level agent pool stabilizes the training process of the high-level router and therefore display a high-profitable performance.

\section{EarnHFT}
\label{append:EarnHFT}
In this section, we introduce the detailed algorithm for market segementing and labeling and the proof of convergence of optimal value supervisor.

\subsection{Algorithm Description}
Here we present a detailed version of the method for labeling the market in Algorithm~\ref{alg:segement_label_market_detail}. This method utilizes a slope and DTW as the criteria to determine whether 2 time series should be labeled as one category.
\label{append:market}
\begin{algorithm}[!htb]
\caption{Market Segmentation \& Labelling}
\label{alg:segement_label_market_detail}
\textbf{Input}: Dataset $\mathcal{D}$ with $N$ timestamp \\
\textbf{Parameter}: risk threshold $\theta$, label number $M$\\
\textbf{Output}: A list of segments with labels\\
\begin{algorithmic}[1] 
\STATE $D' \gets \text{apply low-pass filter to } D$.
\STATE Calculate $\Delta D_t=D_t-D_{t-1}$.
\STATE Find i $\in$ I s.t. $\Delta D_i\Delta D_{i+1}<0$.
\STATE Adding $1$ \& $N$ into I and sort I.
\STATE Initialize $S$ as an empty set.
\FOR {$I_i \in I$}
\STATE do $S_i=\{D_{I_i},D_{I_i+1},...D_{I_{i+1}}\}$ and add $S_i$ to $S$
\ENDFOR

\WHILE{segments in $S$ are not stable}
\FOR{each pair of adjacent segments $s_1, s_2$ in $S$}
\STATE $r_1 \gets \text{ average slope of B\&H net curve in } s_1$
\STATE $r_2 \gets \text{ average slope of B\&H net curve in } s_2$
\STATE $d \gets \text{TWAP between net curve of $s_1$,$s_2$} $
\IF{ $|r_1 - r_2|$ and $|d|$ are small}
\STATE merge $s_1$ and $s_2$ into a single segment in $S$
\ENDIF
\ENDFOR
\ENDWHILE
\STATE Calculate the average slope of buy and hold net curve $r \in R$ \;

\STATE Calculate the risk threshold $H= Q_{1-\frac{\theta}{2}}(R)$, $L= Q_{\frac{\theta}{2}}(R)$ \;

\FOR{i in $\in$ \{ 1,... $|S|$\}}
\STATE $r_i \in R$ is the corresponding slop of segment $s_i$
\IF{$r_i > H$}
\STATE do label $s_i$ as $M$
\ENDIF
\IF{$r_i < L$}
\STATE do label $s_i$ as $1$
\ENDIF
\FOR{$j \gets 2$ to M-1}
\IF{$ L+(j-2)\frac{H-L}{M-2} < r_i\leq L+(j-1)\frac{H-L}{M-2}$}
\STATE do label $s_i$ as $j$
\ENDIF
\ENDFOR
\ENDFOR
\STATE Return the label corresponding to each segment

\end{algorithmic}
\end{algorithm}

\subsection{Proof of Convergence}
\label{append:proof_of_convergence}
Since our method is largely based on q learning, here we give a theoretical proof to show that under a finite MDP, our method can converge to the optimal action value $Q^*$.
Based on proof of Q-learning~\cite{jaakkola1993convergence}, we only need to prove that the operator $H$ is a contraction in the sup-normal where $H$ is defined as follows.
\begin{equation}
\begin{aligned}
&Hq(x,a)\\
&=\lambda\sum_{y \in X}T(y|x,a)(r(x,a,y) + \gamma \argmax_{b} q(y,b))\\&+(1-\lambda)(\frac{1}{|A|}\sum_{a \in A}q(x,a)+Re(x,a))
\end{aligned}
\end{equation}
where 
\begin{equation}
Re(x,a)=q^{*}(x,a)-\frac{1}{|A|}\sum_{a \in A}q^{*}(x,a)
\end{equation}
First, we prove that the optimal Q-function $q^{*}$ is a fixed point of this operator.
\begin{equation}
\begin{aligned}
&Hq^{*}(x,a)=\lambda\sum_{y \in X}T(y|x,a)(r(x,a,y) + \\& \gamma \argmax_{b} q^{*}(y,b))+\\&(1-\lambda)(\frac{1}{|A|}\sum_{a \in A}q^{*}(x,a)+Re(x,a))\\
&=(\lambda)q^{*}(x,a)+(1-\lambda)q^{*}(x,a)\\
&=q^{*}(x,a)
\end{aligned}
\end{equation}
Now we prove operator $H$ is a contraction in the sup-norm
\begin{equation}
  \left|Hq_1-Hq_2\right|_{\inf} \leq k \left|q_1 -q_2\right|_{\inf}
\end{equation}

\begin{equation}
\begin{aligned}
 & \left|Hq_1-Hq_2\right|_{\inf} =
 \\ & \argmax_{x \in X, a \in A}(\lambda \sum_{y \in X}T(y|x,a)\gamma(\argmax_{b_1 \in A} q_1(y,b_1)-
 \\
 &\argmax_{b_2 \in A} q_2(y,b_2))) 
 \\&+\frac{(1-\gamma)}{\left|A\right|}(
  \sum_{a_1 \in A}q_1(x,a_1)- \sum_{a_2 \in A}q_2(x,a_2)
  )\\
 & \leq  \argmax_{x \in X, a \in A} (\lambda \sum_{y \in X}T(y|x,a)\gamma 
 \\&\argmax_{b \in A} (q_1(y,b)-q_2(y,b))\\
 &+(1-\gamma) \argmax_{a_1 \in A}(
 q_1(x,a_1)- q_2(x,a_1))
  )\\
 & \leq  \lambda \gamma \left|\argmax_{b \in A, s \in X}(q_1(s,b)-q_2(s,b))\right|+\\
 &(1-\gamma)\left|\argmax_{b \in A, s \in X}(q_1(s,b)-q_2(s,b))\right|\\
  &=(1-(1-\lambda)\gamma)\left|\argmax_{b \in A, s \in X}(q_1(s,b)-q_2(s,b)) \right|\\
  &=k \left|q_1 -q_2\right|_{\inf}
\end{aligned}
\end{equation}

\section{Experiment Setup}

\subsection{Dataset}
\label{sec:app_dataset}
\(\mathit{BTC/TUSD}\) (BTCT) is currently the rising trading pair in Binance due to the 0 commission fee policy. This trading pair started its life on March 23rd, 2023, and is the most updated dataset. Since it just starts to be traded, the market trend is rather unstable and there is no obvious trend.

\(\mathit{BTC/USDT}\) (BTCU) is the most popular trading pair in Binance due to the extreme stability of USDT. This trading pair owns the most extended trading life, which can date from 2016. The trend of the dataset is relatively stable.

\(\mathit{ETH/USDT}\) (ETH) is one of the most attractive trading pair in Binance due to significant liquidity, volatility, and extremely high correlation with BTC/USDT. This trading pair has drawn considerable attention from radical investors after the boost of Bitcoin. Since the spread of merging ETH, this trading pair decreased by around 20\% in the first half of May, when our test period is presented.

\(\mathit{GALA/USDT}\) (GALA) is a rather niche trading pair in Binance. This trading pair owns one of the highest volatility among all trading pairs and therefore is attractive to radical investors. This trading pair increases around 10\% in the first half of August when our test period is presented.
\subsection{Data Difference}
\label{Appen:app_dataset}
\begin{figure}[!th]
\begin{center}
\vspace{-0.1cm}
\includegraphics[width=0.48\textwidth]{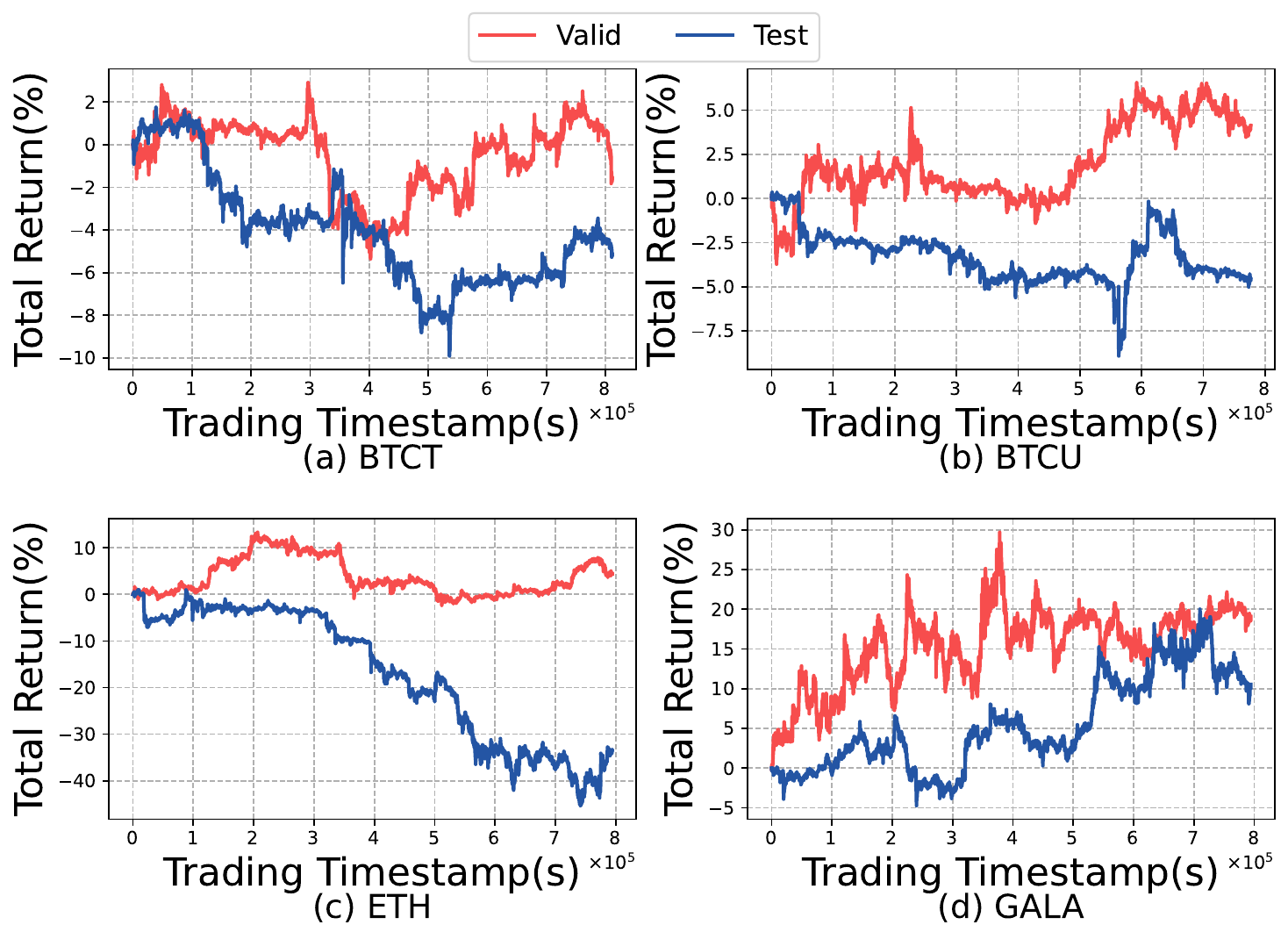}
\end{center}
\caption{Comparison of valid and test dataset}
\vspace{-0.2cm}
\label{fig:Difference_valid_test}
\end{figure}

In this section, we demonstrate the massive difference between valid and test datasets to shown in Figure~\ref{fig:Difference_valid_test}.
In BTCTUSD, there is a stable bear market trend in the test dataset which the valid dataset does not have. There is no huge difference between the valid dataset and test data in BTCUSDT. The test dataset in ETHUSDT is most different from the valid dataset. The trend is stably going up in the valid dataset for ETHUSDT but it drops around 30\% in the test dataset. The test dataset in GALA is more stable than the valid dataset and the overall trends are both going up.

\subsection{Parameter Setting}
\label{appen:parameter_setting}
There are two kinds of parameters during the training: parameters regarding trading, which are identical among different algorithms and different among different datasets, and parameters regarding RL training, which are identical among different datasets and different among different algorithms.

\noindent
\textbf{Trading Setting.}
Here we demonstrate the parameter setting of different datasets in Table~\ref{tab:dataset_trading_setting}.

\begin{table}[!h]
\setlength\tabcolsep{2.8pt}
    \small
  \centering
  \renewcommand{\arraystretch}{1.2}

    \begin{tabular}{|l|cc|}
    \hline
    \textbf{Dataset} & \textbf{Commission Fee (\%)}  & \textbf{Max Holding Coin Number}  \\ \hline
    BTC/TUSD & 0  & 0.01   \\
    BTC/USDT & 0.015  & 0.01    \\
    ETH/USDT & 0.015  & 0.1    \\
    GALA/USDT & 0.015  & 4000    \\ 
    \hline
    \end{tabular}
    \caption{Trading details of the dataset. Here the max holding coin number indicates the maximum position we can hold, which is determined by the unit price of the coin.}
    \label{tab:dataset_trading_setting}
\end{table}
\noindent
\textbf{Training Setting.}
Here we demonstrate the parameter setting of different algorithms. $\alpha$ indicates the Q-teacher's coefficient and the buffer size indicates the replay buffer size in value-based algorithms and batch size in policy-based algorithms in Table~\ref{tab:parameter_setting}. 
\begin{table*}[!th]
\centering
\renewcommand{\arraystretch}{1.2}
  \resizebox{0.95\textwidth}{!}{
  
    \begin{tabular}{|l|ccccccc|}
    \hline
    \textbf{Algorithm} & \textbf{Buffer Size}  & \textbf{Mini Batch Size}  & \textbf{lr}  & \textbf{$\alpha$} & \textbf{Window Length}&  \textbf{Entropy Coefficient} &\textbf{$\tau$} \\ \hline
    EarnHFT & $10^{6}$  & 512 & 5e-4 &128 &- &-&0.005\\
    DDQN & $10^{6}$  & 512 & 5e-4 &-&-&-&0.005 \\
    CDQNRP & $10^{6}$  & 512 & 5e-4 &- &3600&- &-\\
    PPO &512 & 64  & 1e-7 &-  &-&0.01&-\\
    DRA &512  & 64 & 1e-7 &-  &600&0.01&-\\ 
    \hline
    \end{tabular}
}
\caption{Training parameter setting. Here we use the soft update for the target network for value-based methods except for CDQNRP because the random perturbation of the frequency to hard update the target network is one of its contributions. }
\label{tab:parameter_setting}
\end{table*}

\subsection{Experiments Results}
Here we demonstrate some trading processes of EarnHFT and the net value curve for all the datasets and baselines. Here we demonstrate the rationality of the EarnHFT's trading behavior. In BTCU and BTCT, since the unit price of the Bitcoin is larger, the trading cost is also higher, therefore EarnHFT trades less frequently and does not close its position until the price fluctuation has surpassed the trading cost. For ETH, since the trend of the market is changing rapidly, therefore EarnHFT trades much more frequently. GALA remains to go up and there are no apparent fluctuations during the trend therefore, EarnHFT holds it for a long time. In conclusion, in a bull market, the trading frequency is determined by the stability of the market and normally EarnHFT trades less frequently while in a bear market, the agent will trade extremely frequently to seize the profitable trading chances 
\begin{figure}[!th]
\begin{center}
\vspace{-0.1cm}
\includegraphics[width=0.48\textwidth]{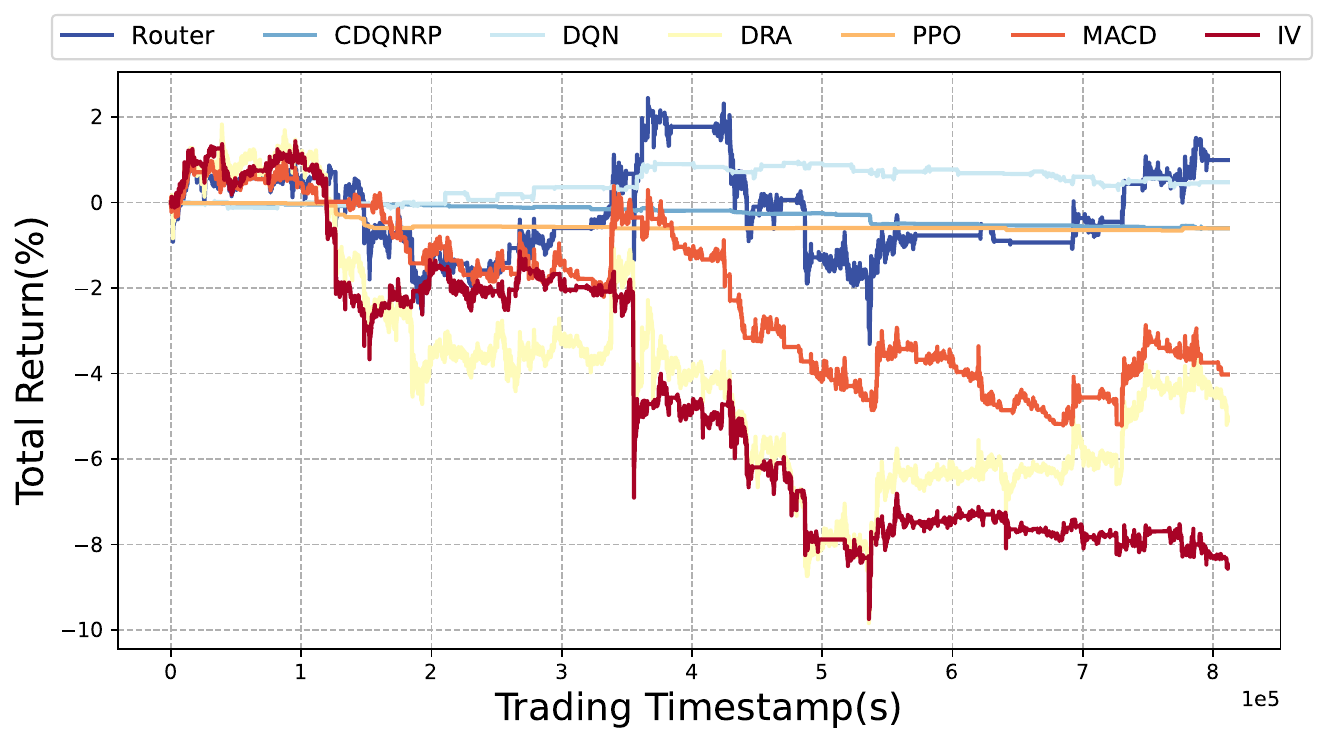}
\end{center}
\caption{The net value curve for BTCT}
\vspace{-0.2cm}
\label{fig:BTCT_net_curve}
\end{figure}

\begin{figure}[!th]
\begin{center}
\vspace{-0.1cm}
\includegraphics[width=0.48\textwidth]{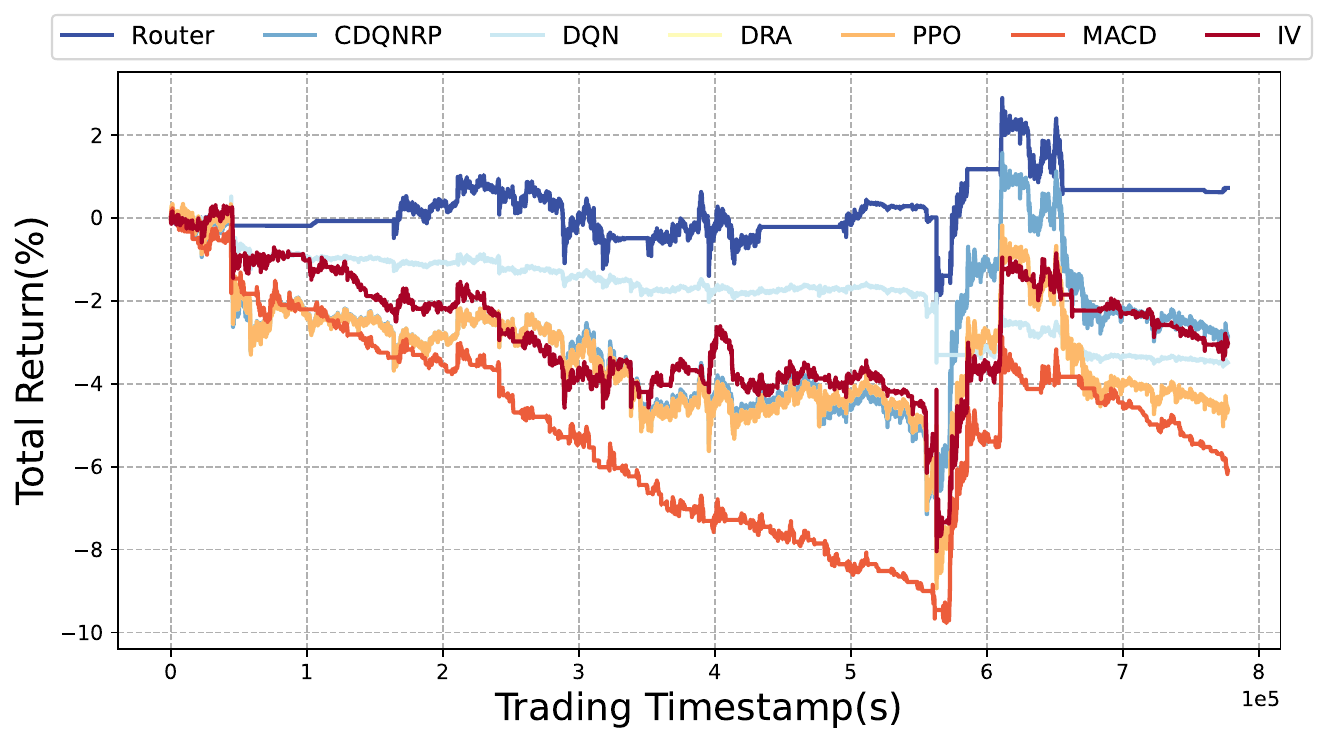}
\end{center}
\caption{The net value curve for BTCU}
\vspace{-0.2cm}
\label{fig:BTCU_net_curve}
\end{figure}

\begin{figure}[!th]
\begin{center}
\vspace{-0.1cm}
\includegraphics[width=0.48\textwidth]{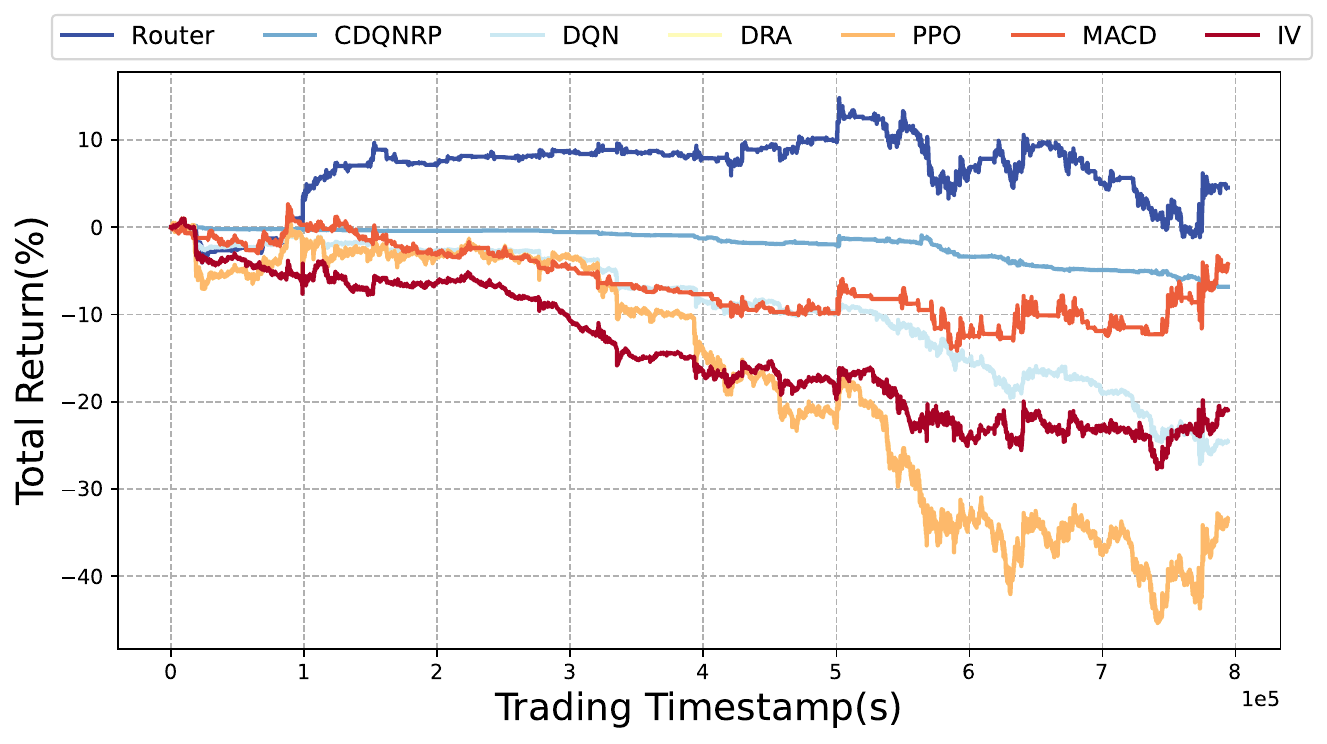}
\end{center}
\caption{The net value curve for ETH}
\vspace{-0.2cm}
\label{fig:ETH_net_curve}
\end{figure}

\begin{figure}[!th]
\begin{center}
\vspace{-0.1cm}
\includegraphics[width=0.48\textwidth]{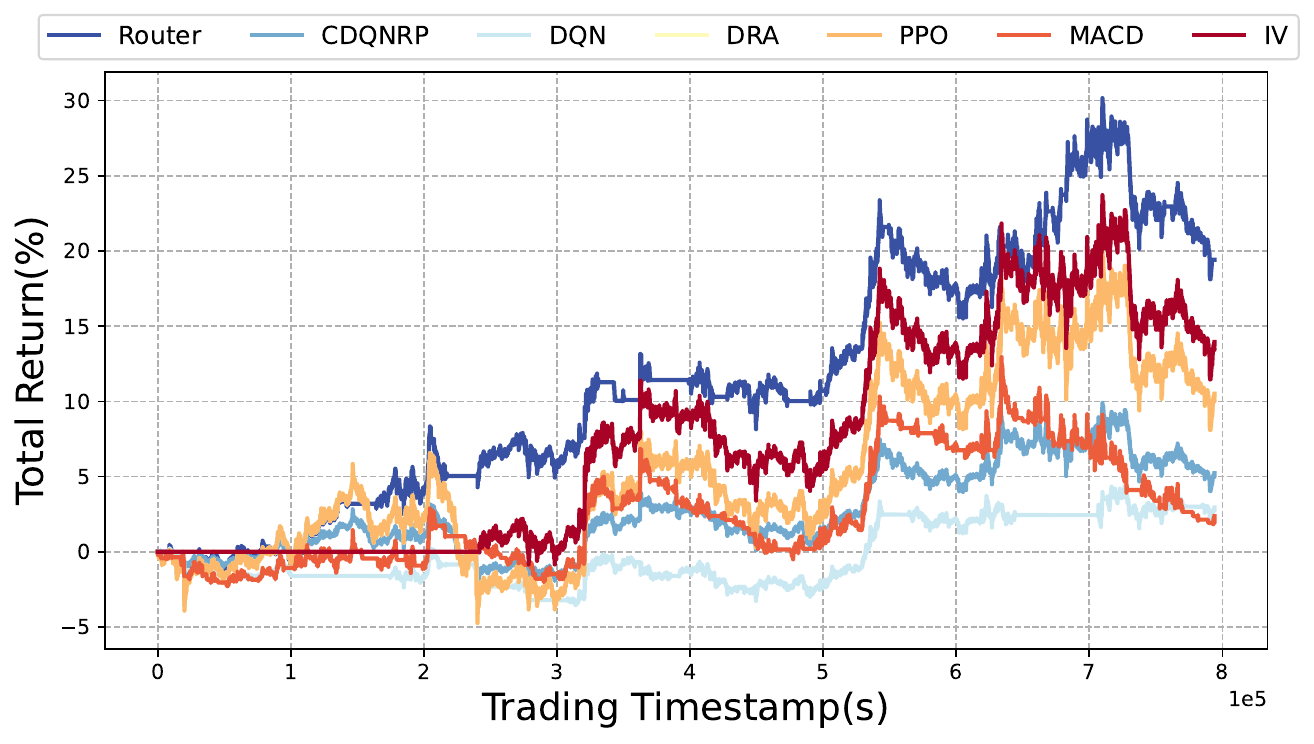}
\end{center}
\caption{The net value curve for GALA}
\vspace{-0.2cm}
\label{fig:GALA_net_curve}
\end{figure}

\begin{figure}[!th]
\begin{center}
\vspace{-0.1cm}
\includegraphics[width=0.48\textwidth]{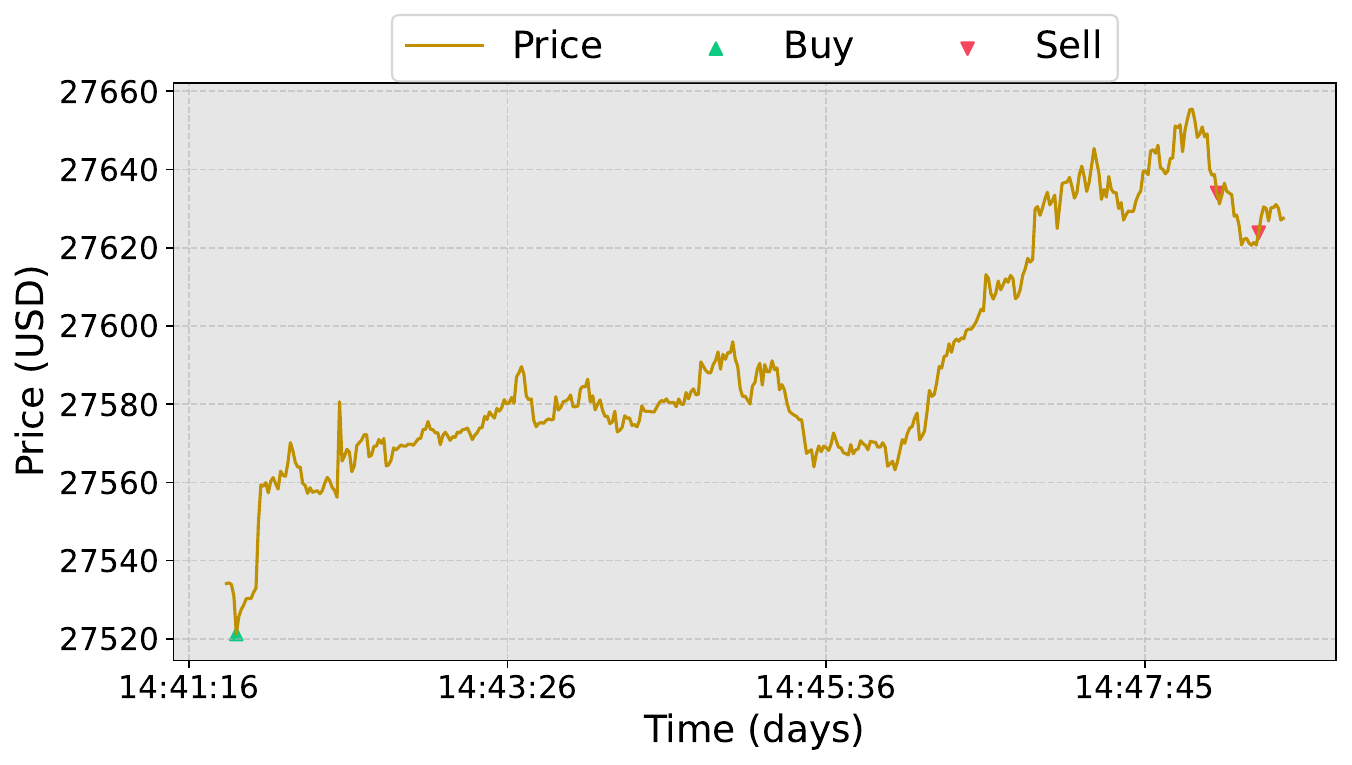}
\end{center}
\caption{The trading example for BTCT}
\vspace{-0.2cm}
\label{fig:BTCT_trade}
\end{figure}

\begin{figure}[!th]
\begin{center}
\vspace{-0.1cm}
\includegraphics[width=0.48\textwidth]{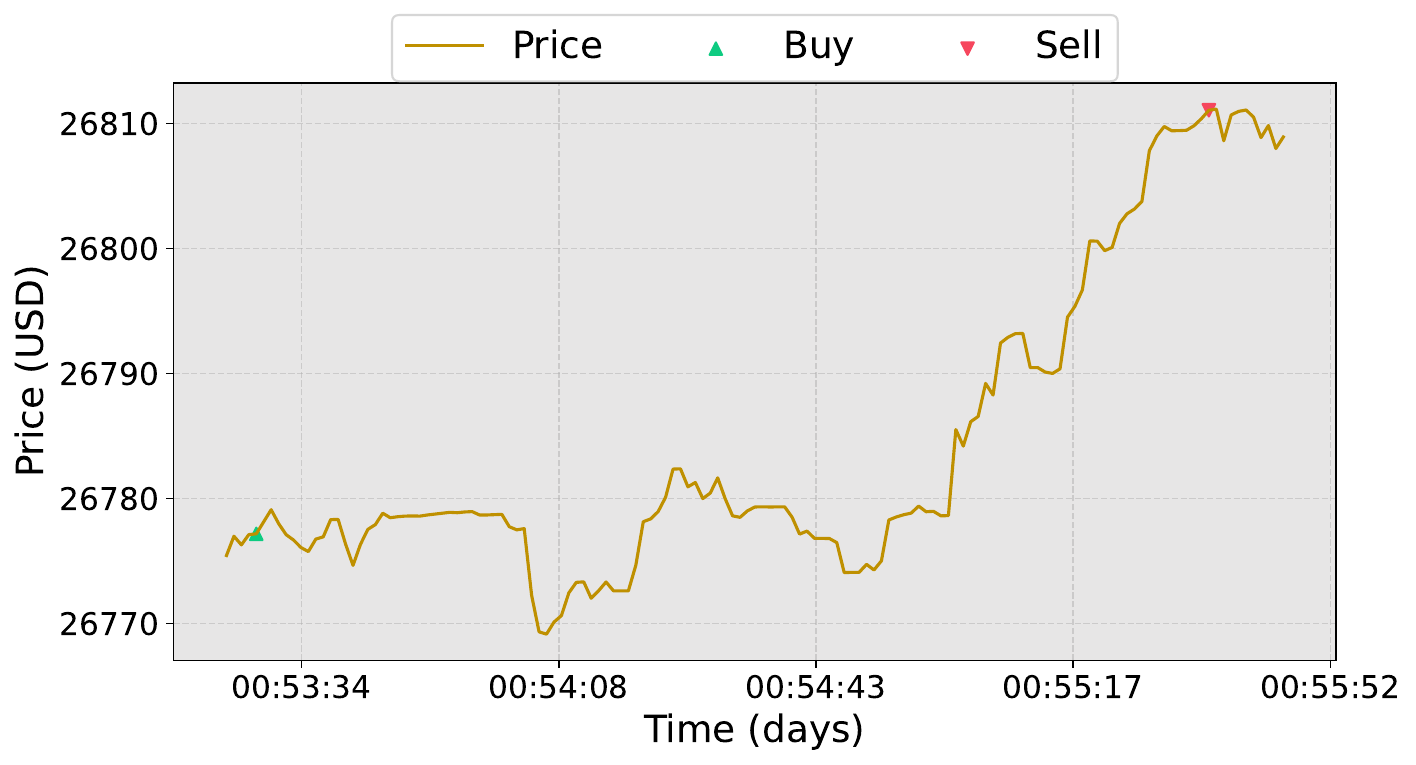}
\end{center}
\caption{The trading example for BTCU}
\vspace{-0.2cm}
\label{fig:BTCU_trade}
\end{figure}

\begin{figure}[!th]
\begin{center}
\vspace{-0.1cm}
\includegraphics[width=0.48\textwidth]{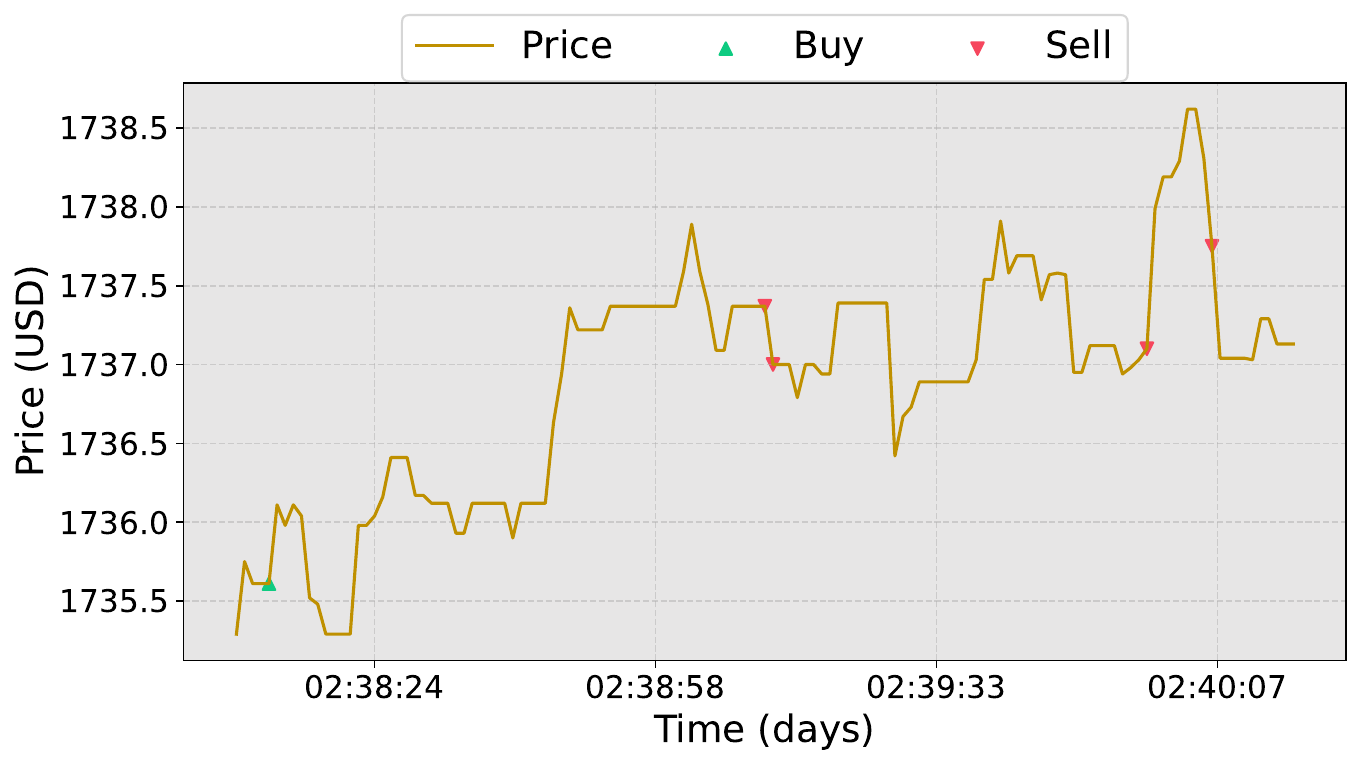}
\end{center}
\caption{The trading example for ETH}
\vspace{-0.2cm}
\label{fig:ETH_trade}
\end{figure}

\begin{figure}[!th]
\begin{center}
\vspace{-0.1cm}
\includegraphics[width=0.48\textwidth]{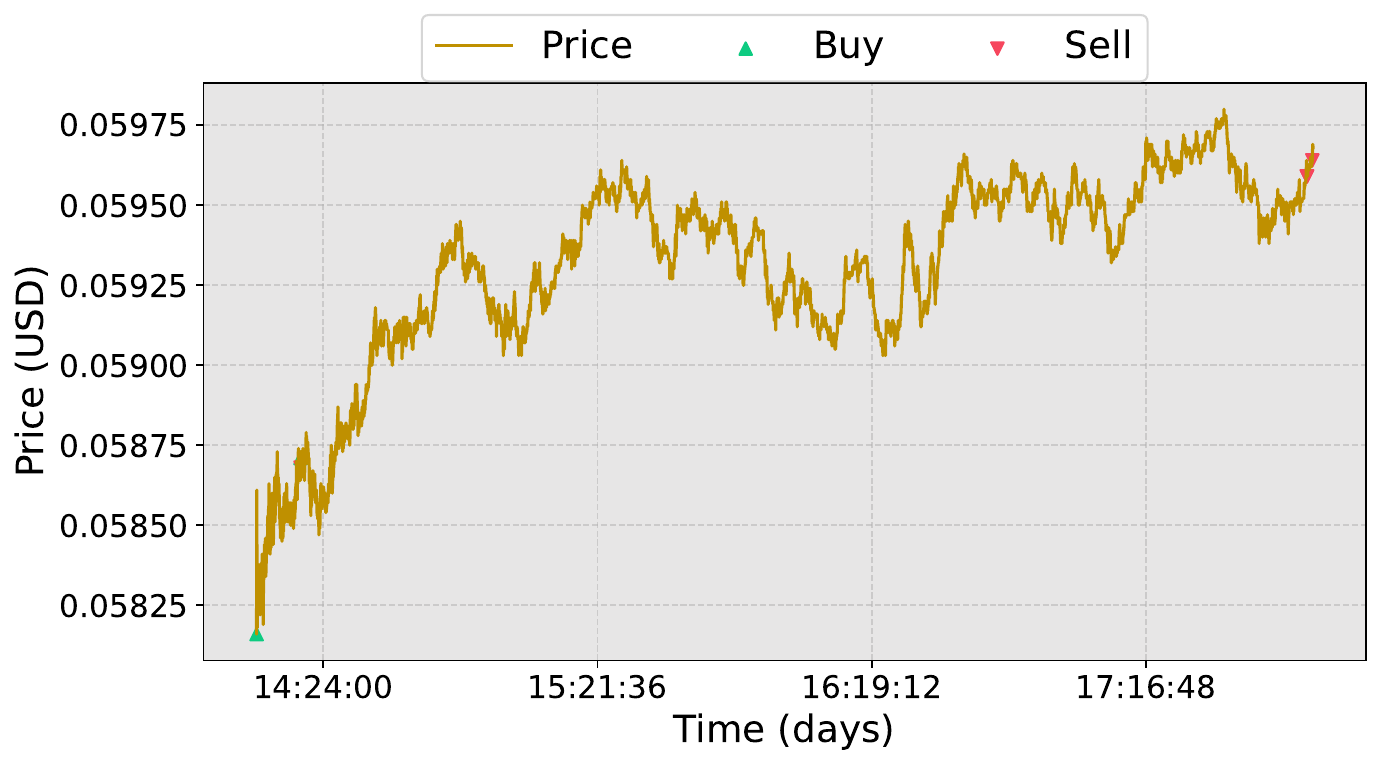}
\end{center}
\caption{The trading example for GALA}
\vspace{-0.2cm}
\label{fig:GALA_trade}
\end{figure}

\bibliography{fin}